\documentclass[12pt]{article} 
\usepackage[T2A]{fontenc}
\usepackage{xcolor}
\usepackage{epsf}
\usepackage[utf8]{inputenc}
\usepackage[russian,english]{babel}
\usepackage{amsmath}
\usepackage{amsfonts}
\usepackage{amssymb}
\usepackage{graphicx}
\usepackage{float}
\usepackage{changes}
\usepackage{comment}
\usepackage{authblk}
\def\aj{AJ}
\def\apj{ApJ}
\def\apjl{ApJL}
\def\apjs{ApJ Suppl. Ser.}

\def\aap{A\&A}

\def\mnras{MNRAS}
\def\nat{Nature}

\def\sovast{Sov. Astron.}

\def\prtphs{Progr. Theoretical Phys. Suppl.}

\def\lum{\rm erg~s$^{-1}$}

\def\beq#1{\begin{equation}\label{#1}}
\def\eeq{\end{equation}}
\def\beqa#1{\begin{eqnarray}\label{#1}}
\def\eeqa{\end{eqnarray}}

\def\myfrac#1#2{\left(\frac{#1}{#2}\right)}
\def\comment#1{\relax}

\def\apgt{\ {\raise-.5ex\hbox{$\buildrel>\over\sim$}}\ }
\def\aplt{\ {\raise-.5ex\hbox{$\buildrel<\over\sim$}}\ }
\newcommand{\ms}{$M_\odot$}
\newcommand{\msun}{$M_\odot$}
\newcommand{\ls}{\mbox {$L_{\odot}$}}

\newcommand{\porb}{\mbox {$P_{\mathrm orb}$}}

\renewcommand\baselinestretch{1.35}

\title{Population synthesis of ultraluminous X-ray sources with magnetised neutron stars  }

\author[1]{A.G. Kuranov \thanks{alexandre.kuranov@gmail.com}}
\author[1,2,3]{K.A. Postnov}
\author[4]{L.R. Yungelson}
\affil[1]{Sternberg Astronomical Institute, M.V. Lomonosov Moscow University}
\affil[2]{A.I. Alikhanov Institute of Theoretical and Experimental Physics}
\affil[3]{Kazan  Federal University}
\affil[4]{Institute of Astronomy, Russian Academy of Sciences}%

\begin{document}
\maketitle




\begin{abstract}
A model of population of ultraluminous X-ray sources with magnetised neutron stars (NULX) in a spiral galaxy with the star formation history
similar to that in the thin disc of the Milky Way is computed using a hybrid approach.  First, applying analytical approximations (code BSE) we construct the ensemble of close binaries (CBS) which can be potential precursors of NULX. Next, evolution with accretion onto magnetised neutron stars (NS) is computed by the evolutionary code MESA. Accretion rate onto NS and X-ray luminosity are calculated for the models of sub- and supercritical discs and for the discs with advection.  
During accretion onto magnetised NS, super-Eddington luminosity $L_\mathrm{X}>10^{38}$ erg~s$^{-1}$ is attained already at the subcritical stage, when the energy release at the inner boundary of the disc defined by the NS magnetosphere is sub-Eddington.  
It is shown that standard evolution of CBS with an account of the peculiarities specific for accretion onto magnetised NS  allows us 
to explain quantitatively observed characteristics of NULX  (X-ray luminosities, NS spin periods, orbital periods and masses of visual components) without additional model assumptions
on the collimation of X-ray emission from NS with high observed super-Eddington luminosity. In a model galaxy with star formation rate 3--5 $M_\odot {\rm yr^{-1}}$ there can exist several NULX. Discovery of a powerful wind from NULX with $L_{\mathrm X} \sim 10^{41}$ erg~s$^{-1}$ may be a signature of super-Eddington accretion onto magnetised NS.

\end{abstract}
	
\onecolumn

\section*{Introduction}
\label{s:intro}

Ultraluminous X-ray sources (henceforth -- ULX) are point-like X-ray sources with equivalent isotropic luminosity in the  0.3–10 keV range exceeding $10^{39}$\,erg~s$^{-1}$. ULX are observed, as a rule, in external galaxies. In the Milky Way only one transient ULX -- 
Swift J0243.6+6124 
(Kennea et al. 2017) is known (see Table 1). ULX were discovered by  \textit{Einstein} in the beginning of the 1980s (Long and van Speybroeck 
1983). ULX are observed both in spiral and irregular galaxies, as well as in the ellipticals. The most luminous ULX are encountered in the star-forming galaxies. ULX are rare: in the local Universe  ($\lesssim$ 40\,Mpc) there are about two candidate objects per galaxy (irrespective of its type), see Walton et al. (2011), Earnshow et al. (2019), Kovlakas et al. (2020). Most likely, this results from the superposition of star-formation effects, stellar evolution, and observational selection. Statistical properties of ULX and their relation to the star formation rate are summarised by Sazonov and Khabibullin (2017).

For a long time, it was deemed that accretors in ULX are intermediate-mass black holes accreting at sub-Eddington rates or stellar-mass black holes 
($M\apgt$10\,\msun) accreting at super-Eddington rates (Mushotzky et. al. 2004).
Bachetti et al. (2014) discovered the fist pulsing ULX. This allowed to identify its accretor with a neutron star. 
Moreover, Brightman et al. (2018) identified with a neutron star the accretor in the source M51~ULX-8,
thanks to the discovery of a resonant cyclotron line, typical for X-ray pulsars, though the pulses themselves were not observed (as yet). 
The analysis of X-ray spectra of 18 NULX (Koliopanos et al. 2017) also suggested that the  majority of ULX may be accreting neutron stars with magnetic fields $\apgt 10^{12}$\,G.

Luminosity of ULX is variable. Quite conditionally, ULXs can be separated into ``quasi-stationary'' or {\it persistent} ones in which 
$L_\mathrm{X}$ throughout observations varied by a factor of several and {\it transient} ones in which an X-ray outburst was observed during which $L_{\mathrm X}$ exceeded  $10^{39}$\,erg~s$^{-1}$ (as a rule, they reside in Be/X binaries). 
At the time of writing, six persistent and four transients were known (see Table 1), for which X-ray luminosity $L_\mathrm{X}$, binary orbital period $P_\mathrm{\rm orb}$, the neutron star spin period $P^*$, and mass of the visual counterpart $M_2$ were estimated. The knowledge of these parameters enables a detailed comparison with the results of population synthesis.

The models of a population of sources with neutron star accretors (henceforth, NULX) have been actively studied and were published earlier 
by Shao and Li (2015), Wiktorowicz et al. (2017, 2019), Marchant et al (2017), Misra et al. (2020).
It was shown in these studies, that within certain model assumptions about the character of supercritical flow and characteristics of radiation of accreting compact objects in close binaries (CBS) it is possible to construct the models of a population of NULX with features similar to observations. It was assumed, by similarity with supercritical accretion discs around black holes, that during accretion onto neutron stars a geometrical beaming of the radiation from the inner regions of accretion discs of NULX occurs.   
As a result, an observer inside the emission cone would overestimate the equivalent isotropic emission of the source (see, e.g., discussion in Wiktorowicz et al. (2017). The influence of the magnetic field of neutron stars on the character of the source emission was not fully addressed.   

In the present paper, we model population  of NULX in CBS with magnetised neutron stars in the sub- and supercritical disc accretion regime. As physical models, we consider the ``standard'' disc model of Shakura and Sunyaev (1973) as applied to magnetised NS by Chashkina et al. (2017), and the model of a supercritical disc with advection around magnetised NS (Chashkina et al. 2019).       
Population synthesis of NULX is accomplished in two steps. First, we use a modified BSE code (Hurley et al. 2002) to separate the range of parameters of CBS that in the course of their evolution may reach the stages of sub- and supercritical accretion.
Next, the mass transfer onto NS is treated using a precomputed grid of models calculated by the code MESA (Paxton 2011). 
For  evaluation of the number of the sources in the galaxies similar to the Milky Way we use star formation history in the Galactic thin disc 
and normalisation to the total mass of stars in the thin disc. 

Results of the computations show that within adopted physical models of accretion, parameters of magnetised NS formation, and the treatment of mass transfer after the Roche lobe overflow by the optical star at the sub- and supercritical accretion stages,
it is possible to explain  successfully both expected number of NULX (per galaxy) and the location of observed NULX in the diagrams
``spin-period of the neutron star -- X-ray luminosity'' ($P^*-L_\mathrm{X}$), ``luminosity -- orbital period''
 ($L_\mathrm{X} - P_\mathrm{orb}$), and $P^*-P_\mathrm{orb}$.
  
\begin{table}[t]
\vspace{6mm}
\begin{center}
\caption{
Parameters of observed ultraluminous X-ray sources with NS (NULX). Asterisks by the numbers in the first column mark transient sources.}

\label{tab:nulx} 
\vspace{5mm}
\begin{tabular}{l|c|c|c|c|c|c|l} \hline\hline
 N & Source   & $L_{\rm min}$      & $L_{\rm max}$     & $P^*$   & $P_{\rm orb}$   & $M_{\rm 2}$ & Ref. \\ 
               && ($10^{39}$ \lum)  & ($10^{39}$ \lum) &  (s)    & (day)          & (\ms)       &       \\
\hline
1 & M82 ULX-2  & $6.6\pm 0.3$        & 18                & 1.37    & 2.53            & $>5.2$      & [1]     \\
2 & NGC7793 P13 & 2.1               & $5\pm0.5$         & 0.415  & $63.9^{+0.5}_{-0.6}$ & 18 - 23       & [2, 3]   \\
3 & NGC5907 ULX-1 & 6.4             & $220\pm30$        & 1.43 - 1.14 & $5.3^{+2}_{-0.9}$ & 2 - 6 & [10] \\
4 & M51 ULX-7     & $\leq 0.3$       & 10  & 2.8        & 2         & $8 - 13$     & [14, 15] \\
5 & M51 ULX-8     & 2              & 20                &            &      8 - 400         &              & [16, 17]  \\
6 & NGC1313 X-2  & 14.4            &19.9               & 1.5         & $\leq 3$    &  $\aplt 12$              & [11, 12] \\
\hline
7$^*$ & NGC300 ULX-1 & 0.6              & 4.7               & 31.6    & $0.8 - 2.1$ yr          & $8 - 10$   & [4, 5] \\
8$^*$ & Swift J0243   & 0               &1.2 -- 2.6          & 9.86       & 27.59       &             & [13] \\ 
9$^*$ & SMC X-3    & 0.2                & 2.5               & 7.8     & 44.86           & $\apgt 3.7$ & [6-8] \\
10$^*$ & NGC2403 ULX-1 & $\aplt 0.001$  & 1.2               & 17.57   & 60 -    100        &               & [9] \\
\hline
\hline
\end{tabular}
\end{center}
\vspace{3mm}
[1] -- Bachetti et al. (2014), 
[2] -- Israel et al. (2017a), 
[3] --  Motch et al. (2014), 
[4] -- Carpano et al. (2018), 
[5] -- Heida et al. (2019), 
[6] -- Tsygankov et al. (2017), 
[7] -- Cowley  and Schmidtke (2004), 
[8] -- Corbet et al. (2003), 
[9] -- Trudolyubov et. al. (2007), 
[10] -- Israel et al. (2017b), 
[11] -- Gris{\'e} et al. (2008), 
[12] -- Sathyaprakash et al. (2019), 
[13] -- Zhang et al. (2019), 
[14] -- Rodr{\'\i}guez Castillo et al. (2020), 
[15] -- Vasilopoulos et al. (2020), 
[16] -- Brightman et al. (2020), 
[17] -- Middleton et al. (2019). 
\end{table}

\section*{Accretion onto magnetized neutron stars}
\label{s:SA}
It is well known that the main difference of the disc accretion onto magnetised NS in close binaries with optical components overflowing Roche lobes from the ``classical'' model of accretion onto black holes (Shakura and Sunyaev 1974) is due to the existence of the NS magnetosphere. At the magnetospheric boundary, the character of accretion flow changes (this process is discussed in detail by Lipunov (1987)).  For sufficiently slow or moderate accretion rates, all matter passing through the disc, after interaction with the magnetosphere, falls onto NS. The main energy release observed in X-ray occurs close to
the surface of NS. If the accretion rate exceeds a certain limit, accretion may become supercritical. This happens if the local energy release
at the inner radius of the disc, bounded by the magnetosphere,  exceeds the Eddington luminosity
  $L_\mathrm{Edd}\approx 1.5\times10^{38} (M_{NS}/M_\odot)$\,\lum. To this luminosity corresponds accretion rate 
$\dot M_\mathrm{Edd}\approx 1.5\times 10^{18}$ g\,s$^{-1}$.

In this case, the basic idea of supercritical accretion, put forward in the pioneering study of
Shakura and Sunyaev (1973), about the outflow of matter inside radius where local energy release begins to exceed the Eddington limit
(so-called spherisation radius $R_s$), is modified (Lipunov 1982; King et al. 2017; Grebenev 2017). Eddington energy release in the disc at the NS magnetosphere boundary
changes the standard expression for the magnetosphere radius (Alf\'{v}en radius) 
$R_A\sim (\mu^2/\dot M)^{2/7}$ (where $\mu$ is magnetic momentum of the NS,  $\dot M$ -- accretion rate). To the first approximation, the radius of magnetosphere ceases to depend on the accretion rate, but is dependent on magnetic field of NS only:  $R_A\sim \mu^{4/9}$.      

The critical accretion rate for which the luminosity at the magnetosphere attains the Eddington limit $\dot M_\mathrm{cr}$, is defined by the equality of the spherisation radius and the magnetosphere radius $R_{s} = R_A$. For the typical NS mass of 
1.4\,$M_\odot$ (which we will apply below for all numerical estimates) 
\begin{equation}
    \label{e:Mcr}
    \dot M_\mathrm{cr}\approx 3\times 10^{19}[\hbox{\text{g\,s}}^{-1}]\mu_{30}^{4/9}\approx 30 \dot M_{-8}\mu_{30}^{4/9}\,.
\end{equation}
Here and below, NS magnetic moment and the accretion rate are expressed in units of 
$\mu_{30}\equiv\mu/10^{30}\,{\rm G\,cm^3}$ and 
$\dot M_{-8}\equiv\dot M/10^{-8} M_\odot\,{\rm yr^{-1}}$, respectively. 
Therefore, 
during disc accretion onto a magnetised NS, the magnetospheric radius can be written as
 
\begin{equation}
\label{r_a} 
R_A=\left(\frac{\mu^2}{\dot M \sqrt{2GM}}\right)^{2/7} \simeq \\
1.6\times10^8[\hbox{cm}]\mu^{4/7}_{30}\dot M^{-2/7}_{-8}, \quad
\dot M < \dot M_\mathrm{cr},
\end{equation}
\begin{equation}
\label{r_sa} 
R_A=\left(\frac{\mu^4 GM}{2L^2_\mathrm{Edd}}\right)^{1/9} 
\simeq 6.5\times 10^7[\hbox{cm}]\mu^{4/9}_{30}, \quad
\dot M > \dot M_\mathrm{cr}.
\end{equation}

We will assume the NS spin period $P^*$ to be close to the equilibrium value,
$P_\mathrm{eq}$, which is defined from the equality of the Alf\'{v}en radius
and corotation radius $R_{co} =(GMP^{2}/4\pi^2)^{1/3}$:
\begin{equation}
\label{e:peqa}
P_\mathrm{eq} \simeq 0.9[\hbox{s}]\mu^{4/9}_{30} \dot M^{-3/7}_{-8}, \quad
\dot M < \dot M_\mathrm{cr}
\end{equation}
at the subcritical stage and 
\begin{equation}
\label{e:peqsa}
P_\mathrm{eq} \simeq 0.2[\hbox{s}]\mu^{2/3}_{30} , \quad
\dot M > \dot M_\mathrm{cr}
\end{equation}
at the supercritical stage.
Such an approximation is justified for large accretion rates onto NS
$\dot M\gtrsim \dot M_\mathrm{cr}\sim 3\times 10^{-7}M_\odot$ yr$^{-1}$,
since at the subcritical stage, if the magnetospheric radius is $\sim 100$ NS radii, the time it takes for the equilibrium rotation to be attained,
 $\tau_{su}=\omega/\dot \omega\sim 100 [\hbox{yr}](\dot M/\dot M_\mathrm{cr})^{-1}(P/\hbox{1 s})^{-1}(R_A/10^8\hbox{cm})^{-1/2}$,  is shorter than the timescale of the equilibrium period variation,
 $\tau_{eq}=7/3 \tau_{\dot M}$,
where $\tau_{\dot M}=\dot M/(d\dot M/dt)$ is the timescale of the evolutionary changes of the accretion rate 
(see examples of evolutionary tracks in Fig.~\ref{fig:trc_1}). At the supercritical stage, the radius of magnetosphere practically does not change, NS periodically enters the propeller stage interchanged by accretion episodes. This is observed apparently in  
M82~X-2 (Tsygankov et al. 2016). Here the NS spin period remains close to the equilibrium value and does not vary significantly.  

Accretion luminosity of a NS is defined, mainly, by the rate of accretion onto its surface. At the subcritical accretion rates,  ($\dot M<\dot M_\mathrm{cr}$), it is equal
to the accretion rate in the disc, while in the supercritical regime 
($\dot M>\dot M_\mathrm{cr}$) it is limited by the leakage\footnote{It is not excluded
that at very high accretion rates onto NS a fraction of energy may be taken away by neutrinos (Basko and Sunyaev 1976; Mushtukov et al. 2018).} through the magnetosphere, i.e.,  
\begin{equation}
\label{e:lx_na} 
L_\mathrm{X}=0.1 \dot{M} c^2 \simeq L_\mathrm{Edd}\dot M_{-8},~~ \dot M < \dot M_\mathrm{cr},
\end{equation}
\begin{equation}
\label{e:lx_nsa} 
L_\mathrm{X}=L_\mathrm{Edd}\myfrac{R_A}{R_{NS}}
\simeq 
65 L_\mathrm{Edd}\mu^{4/9}_{30},~~ \dot M > \dot M_\mathrm{cr}.
\end{equation}

Schematically, the dependence of the main parameters of a magnetised NS -- its  Alf\'{v}en radius 
$R_A$, equilibrium spin period  $P_\mathrm{eq}$, and X-ray luminosity $L_\mathrm{X}$ on the accretion rate $\dot M$ is shown in Fig.~1 (see also Postnov et al. 2019).
\begin{figure}
\label{f:RPL}
\vskip -2cm
\centering{
\includegraphics[width=0.5\columnwidth]{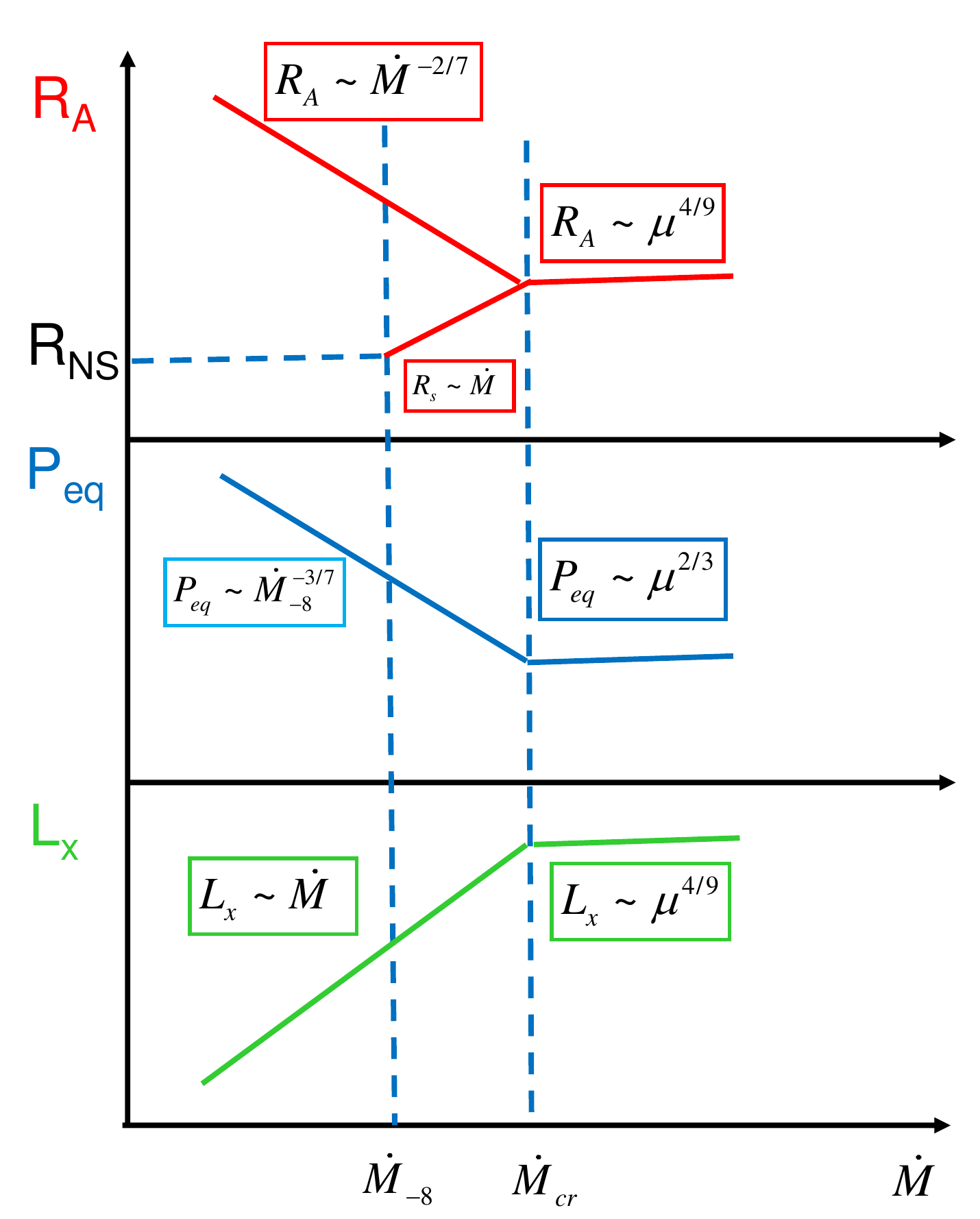}
\caption{Schematic dependence of Alf\'{v}en radius 
$R_A$, spherisation radius $R_s$, equilibrium spin period  $P_\mathrm{eq}$, and X-ray luminosity $L_\mathrm{X}$ of a neutron star with magnetic momentum $\mu$ on accretion rate $\dot M$.}
}
 
\end{figure}

Supercritical accretion onto NS was revisited by Chashkina et al. (2017).
Restructuring of the NS magnetosphere occurs at the critical accretion rate corresponding to the condition of the local Eddington energy release at the inner boundary of the disc, which is equivalent to the increase of semi-thickness of the disc $H$ in the vicinity of the inner disc rim to $H/R\sim 1$ (Eq.~(66) in Chashkina et al.\ (2017)). 
This value is about three times as high as 
$\dot M_\mathrm{cr}$ from Eq.~(\ref{e:Mcr}). However, even this is not the  maximum possible value. As discussed in detail in Chashkina et al. (2019),
at high accretion rate onto NS,  corrections to the accretion disc structure due to advection 
become important because of the increased thickness of the disc  
(see  Lipunova (1999) for the theory of supercritical accretion discs with outflows with an account of advection). 
Then, according to Eq.~(28) from Chashkina et al. (2019), the critical accretion rate onto magnetised NS may increase up to
\begin{equation}
\label{e:mcrad}
    \dot M_{cr,ad}\sim 5\times 10^{20}[\hbox{g\,s}^{-1}]\mu_{30}^{4/9}.
\end{equation}

In the present study, we will use two models: (i) with a conservative value of the critical accretion rate (\ref{e:Mcr}) for the standard accretion discs and (ii) a model with an order-of-magnitude higher critical rate (\ref{e:mcrad}) for advection discs. We will ignore  weak dependence of supercritical Alf\'{v}en radius on the accretion rate for advection discs (less than by a factor two for variation of accretion rate by three orders of magnitude, see Fig.~12 in Chashkina et al (2019)).

It is important to note that  Eq.~(\ref{e:Mcr}), and, especially, Eq.~(\ref{e:mcrad}) show that 
in the case of accretion onto magnetised NS, high X-ray luminosities
$L_\mathrm{X}>10^{39}$ \lum\ can be achieved already at the subcritical accretion stage,
i.e., when the local energy release at the inner rim of the accretion disc is still sub-Eddington and the disc is not blown-up by radiation pressure to $H/R_A\sim 1$.
Thereat, the luminosity in the accretion column of the NS may be significantly 
super-Eddington (see, e.g., Mushtukov et al. (2015)).
   
Possible beaming of the X-ray emission of supercritical accretion discs around magnetised NS should be discussed separately. The beaming factor 
$b$ is defined as the ratio of the observed X-ray luminosity  $L_\mathrm{X}$ and the 
luminosity of an isotropic source computed from the observed flux $F_x$, $L_{iso}=4\pi d^2 F_x$ ($d$ -- the distance to the source), i.e.  
$b\equiv L_{X}/L_{iso}\le 1$. In the case of a source with  radiation into counter-directed cones with half-opening angles  $\theta$,  $b=(1-\cos\theta)$.
This factor is usually taken into account in the modelling of the population of NULX (see, for instance, Wiktorowicz et al. (2017)). However, the factor which is used is based on the extrapolation of beaming of X-ray radiation from supercritical discs around black holes, for which $b\sim \dot M^{-2}$ (King 2009). 
In the case of magnetised NS, as noticed above, the contribution from the energy release of the disc
to the observed X-ray luminosity will be substantially lower than the contribution from the accretion column 
close to the NS surface (see Eq.~(\ref{e:lx_nsa})).
There remains only geometrical beaming of X-ray emission by a thick disc with half-thickness $H/R_A\sim 1$. For NS radiation screened by the thick inner disc, the factor 
$b=(1-H/\sqrt{H^2+R_A^2})=(1-1/\sqrt{2})$ is accounted for in our computations both for accretion and advective supercritical discs. We neglect the X-ray beaming from the accretion column itself, which is averaged by NS rotation.

\section*{The method of calculations}
\label{s:method}
As mentioned in the Introduction, we apply a two-step hybrid method of population synthesis of NULX in close binaries. This method combines rapid simplified computation of the binary evolution up to the RLOF by the optical star in a binary with NS component by the BSE code with subsequent detailed calculation of the mass-transfer stage by evolutionary code MESA. Such a hybrid method was successfully applied, in particular, for modelling populations of SN~Ia precursors, cataclysmic variables (see, e.g., Chen et al. (2014), Goliash and Nelson (2015)), as well as a population of ULX (Shao and Li 2015).

At the first step, using the modified code BSE based on the analytical approximations for the description of the evolution of single and close binary stars, we modelled population of NS in pairs with non-degenerate stars (visual components) that may potentially
become NULX, using $10^7$ massive initial binaries
We used the Salpeter initial mass function ($dN/dM\sim M^{-2.35}$), a flat distribution of 
binary component mass ratios $q=M_2/M_1\le 1$  in the [0.1,1] range and a flat initial orbital eccentricity distribution in the  [0,1] range.
Stellar binarity rate was assumed to be 50\% (i.e., 2/3 of all stars are binary components).
The distribution of initial binaries orbital periods followed Sana et al. (2015): $f(\log \porb) \propto \log \porb^{-0.55}$. 
Stellar winds of massive stars and helium stars were treated using formulas from Vink et al. (2001) and Vink (2017), respectively. 
Evolution of binaries in common envelopes was treated according to Webbink (1984) and de Kool (1990) formalism with parameter $\alpha=1$.
The parameter $\lambda$ describing binding energy of stellar envelopes was taken after Loveridge et al. (2011). 

NSs with mass 1.4\,$M_\odot$ were assumed to be produced by the collapse of iron cores of massive stars. We have used approximate criteria for the NS formation after Giacobbo and Mapelli (2018). The nascent NS obtained a kick following Maxwellian distribution with dispersion $\sigma=265$ km s$^{-1}$ (Hobbs et al. 2005). Rare NSs formed due to the electron captures in O-Ne-MG cores of 8.5-8.8 $M_\odot$ 
stars (Siess and Lebreuilly 2018) that experienced previously mass exchange   
(Miyaji et al.  1980),
were assigned a low kick   
30~km~s$^{-1}$ (this entirely arbitrary value does not influence the results).  
Magnetic moments of NSs followed a log-normal distribution (Faucher-Gigu{\`e}re and Kaspi 2006)
\beq{e:mu}
f(\mu)\sim \exp\left[\frac{(\log\mu - \log\mu_0)^2}{\sigma_\mu^2}\right]\,.
\eeq
We made computations for $\log(\mu_0)$\,[G\,cm$^3$]=30.6\footnote{In the original paper by Faucher-Gigu{\`e}re and Kaspi  $\log\mu_0=30.35$ [G\,cm$^3$] for the NS radius 10\,km. We assumed  $R_{NS}=12$~km.} and $\log(\mu_0)$\,[G\,cm$^3$]=31.6 and 
$\log(\sigma_{\mu})$\,[G\,cm$^3$]=0.55.
Decay of the NS magnetic fields due to accretion was ignored because the timescale of mass transfer in massive CBS is much shorter than the possible timescale of NS magnetic field decay. Nevertheless, the decay of NS magnetic field  may reduce the duration and X-ray luminosity of NULX stage in binaries with  low initial mass of the donors in which the mass transfer stage can last for a long time 
($M \aplt 3$\,\ms, see Appendix).

Our calculations were performed for stars with solar metallicity (Z=0.02). Considering that all known NULX are discovered in nearby spiral galaxies, this assumption also does not change our conclusions substantially.  
To evaluate the number of NULX in a model galaxy, we assumed  star formation rate similar to
the one in the thin disc of the Milky Way (Yu and Jeffery 2010): 
\beq{e:SFR}
\frac{\mathrm{SFR}(t)}{M_\odot\,\mathrm{yr}^{-1}}=\left\{
\begin{array}{lr}
11e^{-\frac{t-t_0}{\tau}}+0.12(t-t_0), & t\ge t_0\\
0, & t<t_0,
\end{array}
\right.
\eeq
where $t$ is the time in Gyr, $t_0$=4~Gyr is the time of the beginning of star formation in the disc, the parameter  $\tau=9$~Gyr. In this model, the current mass of the thin Galactic disc  (at the age of 14 Gyr) is $M_G=7.2\times 10^{10} M_\odot$.

At the second step of the modelling, the evolution of semi-detached CBS with NS components was calculated in detail using the evolutionary code MESA (Paxton et al. 2011, version r12778). 
A grid of tracks was computed for donor masses ranging from 0.75 to 50~\ms. In the   (0.75 - 10)\,\ms\ range, the mass step was $0.25 M_\odot$, between 10 and 50\, \ms\ it was  2\,\ms.  The initial semi-major axes of orbits 
were in the range
$0.9< \log (a_{ini}/R_\odot) < 3.5 $ with the step  $\Delta(\log (a_{ini}/R_\odot)) = 0.1$. 
Next, for every pair of parameters  [$X,Y$], where  $X,Y \equiv \{L_X, P_{orb}, P^*, M_2$\}, the time spent as NULX in the interval  $\Delta t_k$ in the
cell  $[X_i,Y_j]$ was computed and convolved with the probability of formation of the given system (per unit mass), as calculated by BSE:
\beq{e:N}
\frac{\Delta N(t_k,t_k+\Delta t_k)}{\Delta X_i\Delta Y_j\langle{m_{BSE}\rangle}} = 
\frac{\sum\limits_{l=1}^{N_{MESA_{i,j}}}\Delta t^{k,l}_{NULX}}{N_{MESA_{i,j}}\Delta t_k}\times \frac{N_{BSE_{i,j}}}{N_{BSE}}\times\frac{1}{\langle{m_{BSE}\rangle}}\,.
\eeq
Here,  $\Delta t^{k,l}_{NULX}$ is the duration of NULX stage for the system $l$, 
$N_{MESA_{i,j}}$ and $N_{BSE_{i,j}}$ are the numbers of systems in the cell 
$[X_i,Y_j]$ according to the MESA and BSE grids, respectively,  
$N_{BSE} = 10^7$ is the number of initial systems computed by BSE, 
$\langle{m_{BSE}\rangle} \approx 0.44$ \ms\ is the average mass of a binary for the assumed initial mass function and the binary mass ratio distribution.
The distribution obtained for the time instant  $t_k$ is convolved with the star formation history in the model galaxy (\ref{e:SFR}) and is presented for the time 14\,Gyr. 

To describe the orbital evolution of close binaries at the mass-exchange stage, we applied the formalism presented in Soberman et al. (1997). Here the equation relating the 
angular momentum and mass loss from the system through the vicinity of the second Lagrange point $\mathrm{L_2}$ has the form 
\begin{multline}
\label{dm} 
\dot J_{ml} = [(\dot M_{2,w} + \alpha_\mathrm{mt}(dM/dt)_{L1})M^2_1 \\
+ (\dot M_{1,w} + \beta_\mathrm{mt}(dM/dt)_\mathrm{L1})M^2_2]
\times \frac{a^2}{(M_1 + M_2)^2}\frac{2\pi}{P_\mathrm{orb}} \\
+ \gamma_\mathrm{mt}\delta_\mathrm{mt}(dM/dt)_\mathrm{L1}\sqrt{G(M_1 + M_2)a}\,.
\end{multline}
Dimensionless parameters 
$\alpha_\mathrm{mt},~\beta_\mathrm{mt},~\delta_\mathrm{mt}$ 
mean the fractions of the matter that outflew from  the donor via L$1$ and left 
CBS from the vicinity of the donor, 
the accretor,  and the coplanar circumbinary torus with the radius  
$ \gamma^2_\mathrm{mt} a$ (through the vicinity of 
$\mathrm{L_2}$), respectively.
If the stellar wind mass loss is neglected  ($\dot M_{2,w}$=0), the efficiency of mass transfer from the donor to the accretor is 
$f_\mathrm{mt} = \min [1 - \alpha_\mathrm{mt}-\beta_\mathrm{mt}-\delta_\mathrm{mt}, \dot M_\mathrm{Edd}/(dM/dt)_{\rm L1}]$.
Here, $\dot M_\mathrm{Edd}$ is the accretion rate onto NS at which the Eddington luminosity is attained, and the matter outflow caused by radiation pressure begins. 

Observations of SS433 reveal the presence of gas outflow from the system both in the form of a quasi-spherical wind from a supercritical accretion disc and through the vicinity of the 
$\mathrm{L_2}$ point (see review by Cherepashchuk et al. (2020)). The location of the 
$\mathrm{L_2}$
point depends on the binary mass ratio and is characterised  by the value 
$\gamma_\mathrm{mt}^2\approx 1.2$. Thus,  $\gamma_\mathrm{mt}\approx 1.1$ can be taken as the minimum possible value of the parameter describing mass loss via 
$\mathrm{L_2}$ (see Fig.~1 in Cherepashchuk et al. (2018)). 
Recent observations of gas flow around  SS433 by the optical interferometer VLTI GRAVITY (Weisberg 2019) provide an evidence for a much more effective angular momentum loss from the system via a circumbinary disc corresponding to  $\gamma_\mathrm{mt}\sim 5$ (Cherepashchuk et al. 2019).
In our calculations, we accepted $\gamma_\mathrm{mt}$=3.0 as a compromise between its value in $\mathrm{L_2}$ equal to 1.15
and $\gamma_\mathrm{mt} \approx 5.0$ found by Cherepashchuk et al. (2019) for SS433.  

We set in the calculations the following values of dimensionless parameters characterizing the efficiency of mass transfer between components of CBS with visual star ($M_2$) overflowing  Roche lobe: 
$\alpha_\mathrm{mt} = 0.0$, 
$\beta_\mathrm{mt} = 0.0$  at the subcritical accretion stage and 
$\beta_\mathrm{mt}=((dM/dt)_\mathrm{L1}-\dot M_\mathrm{Edd})/(dM/dt)_\mathrm{L1}$ at the supercritical accretion stage.
In MESA,  the value  $\dot M_\mathrm{Edd}\approx 1.5\times 10^{-8}(M/M_\odot)$ \ms\,yr$^{-1}$ is set for compact objects of the mass $M$ without magnetic field. 
In our case, in this formula $\dot M_\mathrm{cr}$ should be used 
instead of $\dot M_\mathrm{Edd}$. This means that the orbital evolution of CBS with mass transfer should be considered separately for each binary, which would be too expensive computationally.  
In the population synthesis calculations we preferred to fix this parameter in MESA, i.e., we used the maximum possible value of
 $\beta_\mathrm{mt}$.   
Note, however, that the orbital evolution is more sensitive to the  parameters 
of non-conservative mass loss $\delta_\mathrm{mt}$ and $\gamma_\mathrm{mt}$ than to 
$\beta_\mathrm{mt}$ which describes the mass loss from the system in the Jeans mode at the supercritical accretion stage (see Appendix for more detail). 

The parameter of non-conservative mass loss due to escape of matter through  Lagrange point  
$\mathrm{L_2}$ was set to $\delta_\mathrm{mt} = 0.1$. 
This value is motivated, in particular, by observations of the change of orbital periods 
of semi-detached binaries (see, e.g., Erdem and  \"Otzt\"urk (2014)).

The magneto-rotational evolution of NS was treated using detailed formulae presented in Lipunov et al. (2009).

After the Roche lobe overflow by the donor, mass transfer rate via the inner Lagrange point $\mathrm{L_1}$,  $(dM/dt)_{\rm L1}$, as a rule, very rapidly increases and starts to 
exceed the critical accretion rate onto NS  $\dot M_\mathrm{cr}$. 
There can be two options for further evolution. Despite the high 
mass transfer rate through $\mathrm{L_1}$
$(dM/dt)_{\rm L1}$,
which in some cases may be $\sim 10^{-2} - 10^{-1}$\,\ms\,yr$^{-1}$,
MESA finds solutions of the system of stellar structure equations for which the star remains confined to the Roche lobe. The evolution ends by the formation of a white dwarf or a helium star depending on the initial mass\footnote{An ULX with Wolf-Rayet optical component is known (Qiu et al. 2019), but the nature of accretor is still not defined.}. If the solution is not found and calculations terminate, this means that the radius of the donor continues to increase, and the system enters the common envelope stage. 
In both cases, we assume that the system is a NULX as long as the X-ray accretion luminosity is above $10^{39}$\,\lum.

\section*{Results of computations}
\label{s:example}
\subsection*{Examples of evolutionary tracks with NULX computed by MESA}

\begin{figure} 
\includegraphics[width=1.0\textwidth]{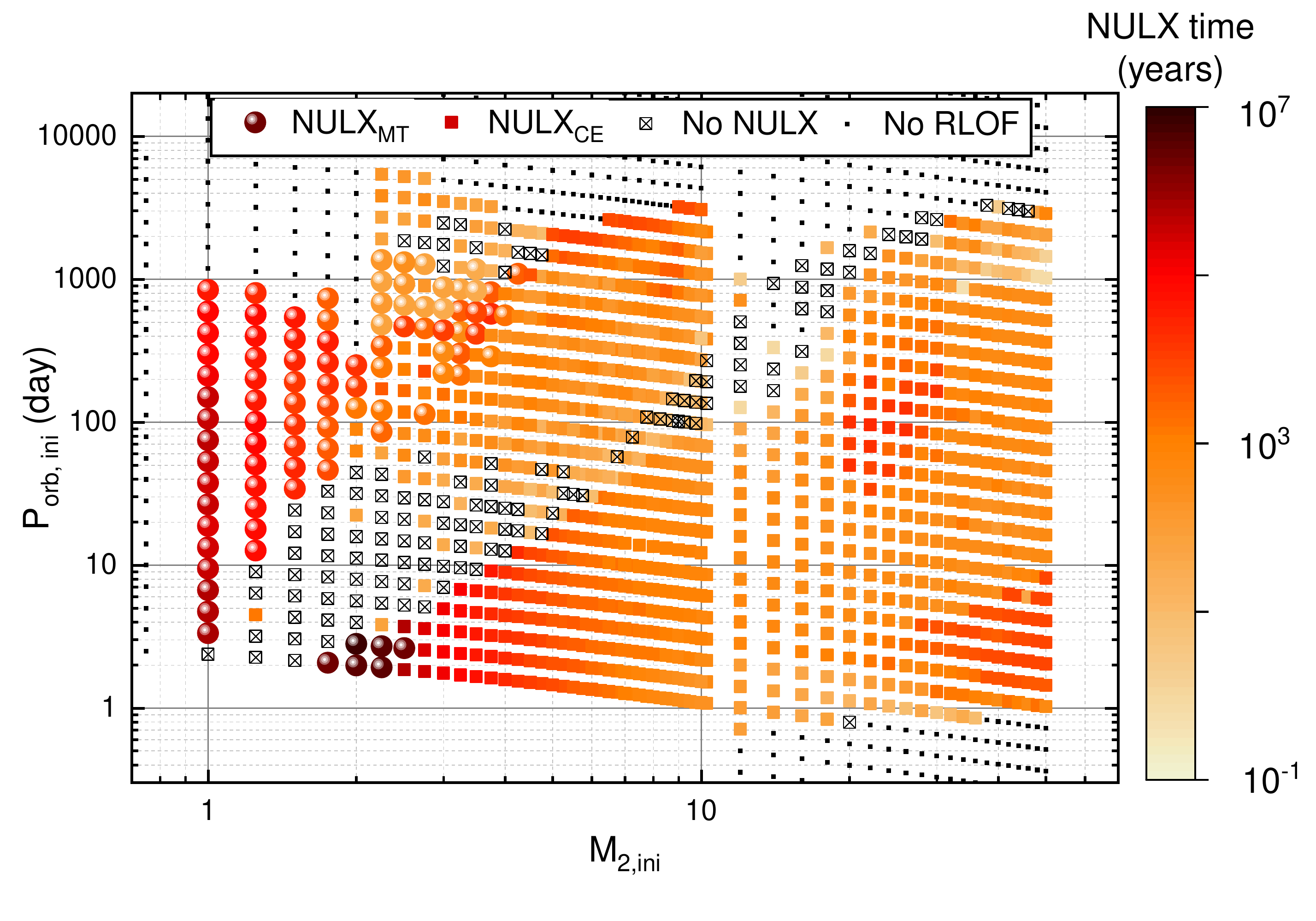}
\caption{Time duration of the stage with the mass-transfer rate exceeding $10^{-7} M_\odot$ yr$^{-1}$ for the grid of stellar models 
with $0.75 -  50 M_\odot$ donors 
and initial orbital semiaxes $0.9< \log (a_\mathrm{ini}/R_\odot) < 3.5 $ computed by MESA, in the coordinates 
$M_{\rm 2,ini} -  P_{\rm orb,ini}$. The subscript {\it ini} refers to the systems with donors at ZAMS accompanied by NS. Dots mark the systems in which donors did not overfill Roche lobes.
Strikethrough squares show the systems in which NULX stage does not occur.  
The filled circles show the systems with stable mass transfer (NULX$_{\rm mt}$).
Squares mark the systems that end their evolution by the common envelope formation (NULX$_{\rm ce}$).
The colour scale reflects the time duration of NULX in years. 
}
\label{fig:mesa_grid}
\end{figure}

\begin{figure} 
\includegraphics[width=0.5\textwidth]{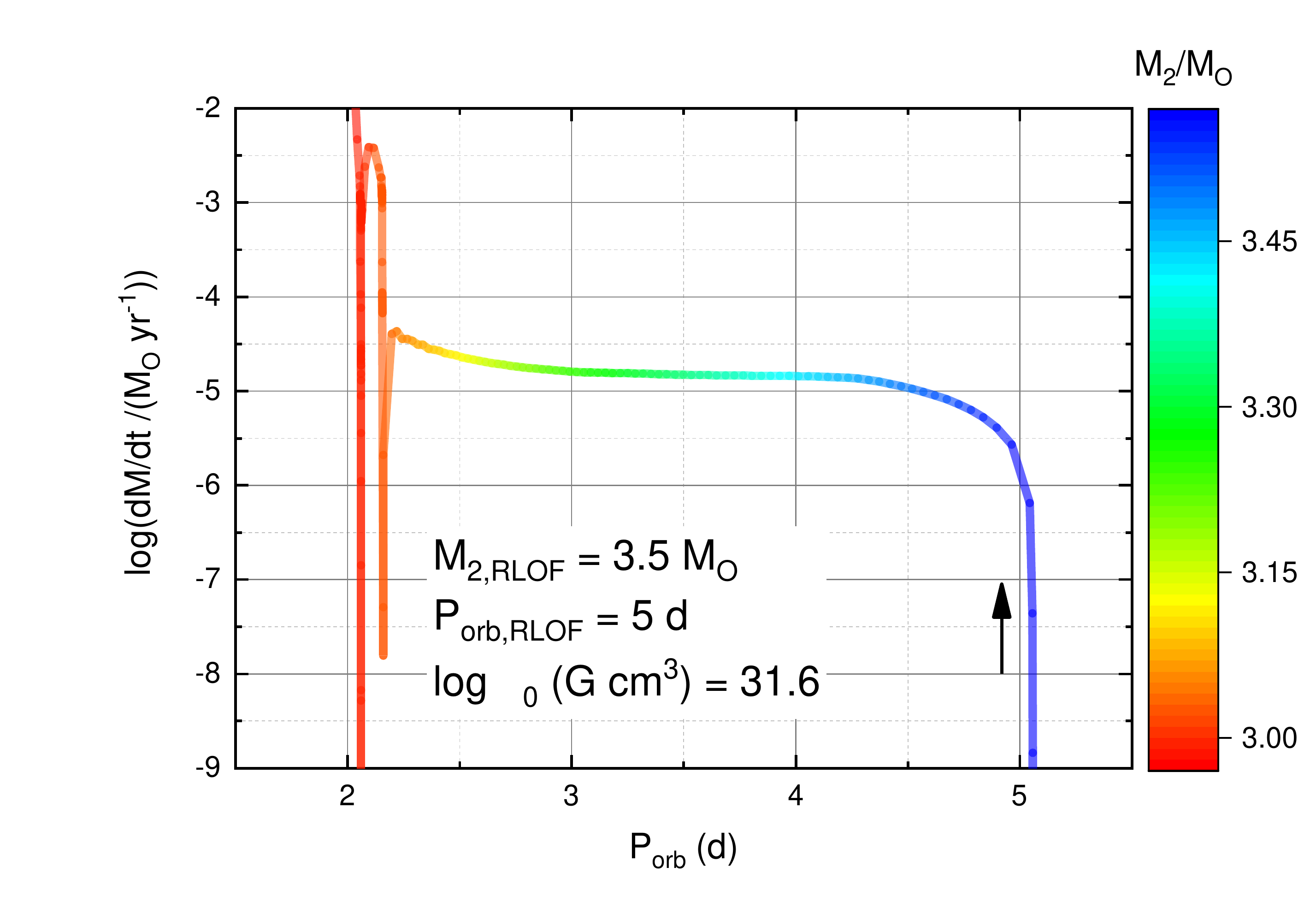}
\includegraphics[width=0.5\textwidth]{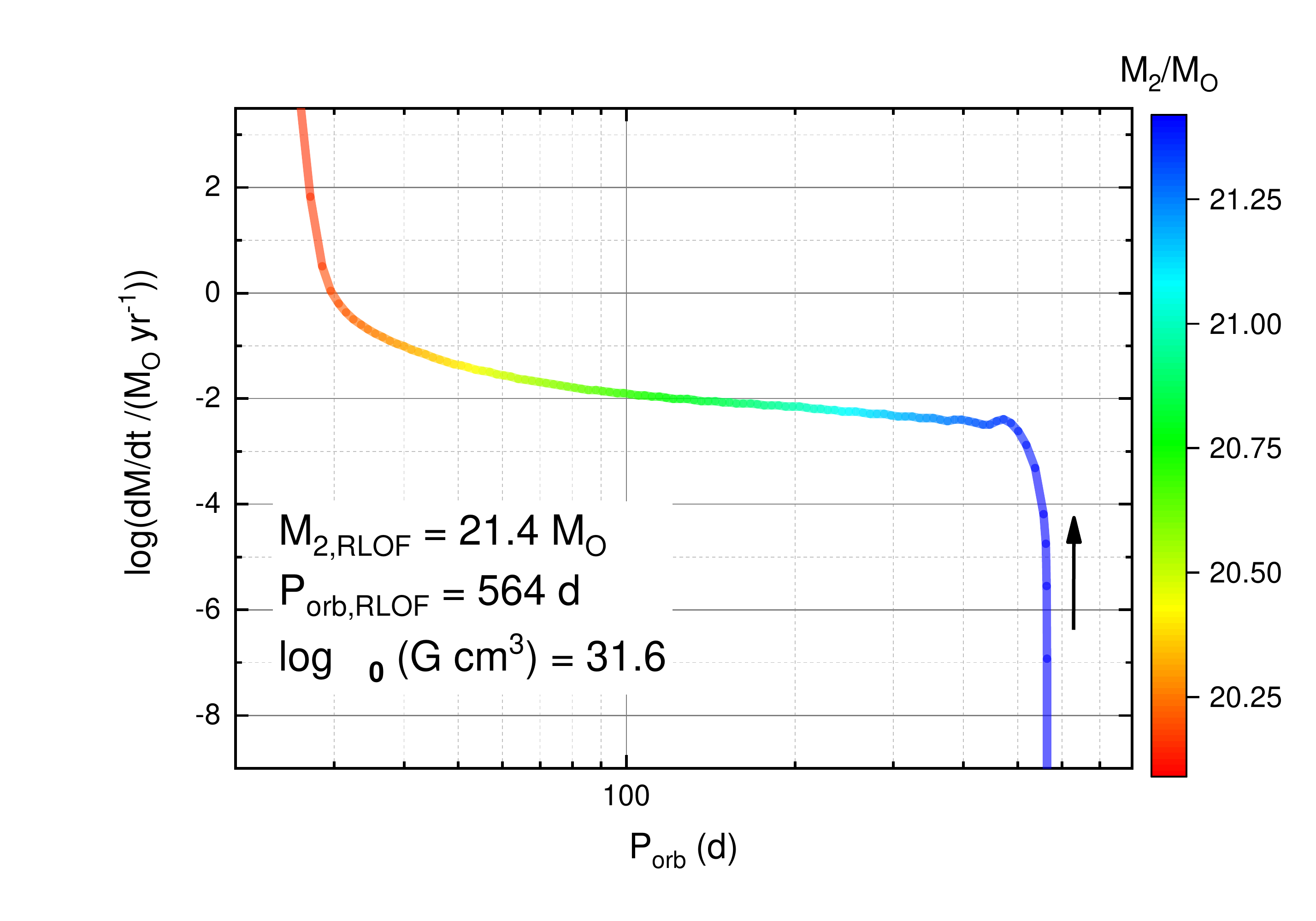}
\vfill
\includegraphics[width=0.5\textwidth]{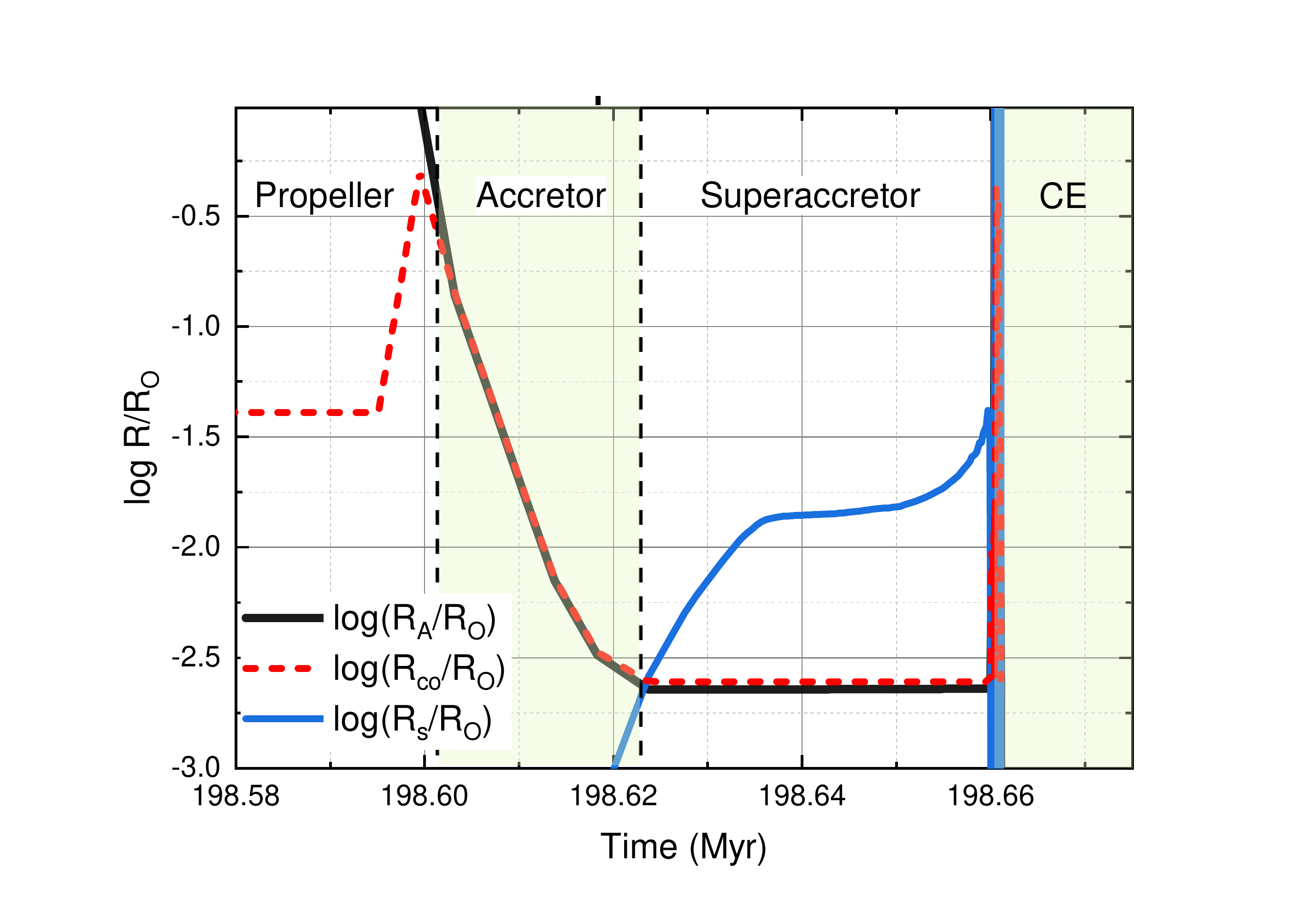}
\includegraphics[width=0.5\textwidth]{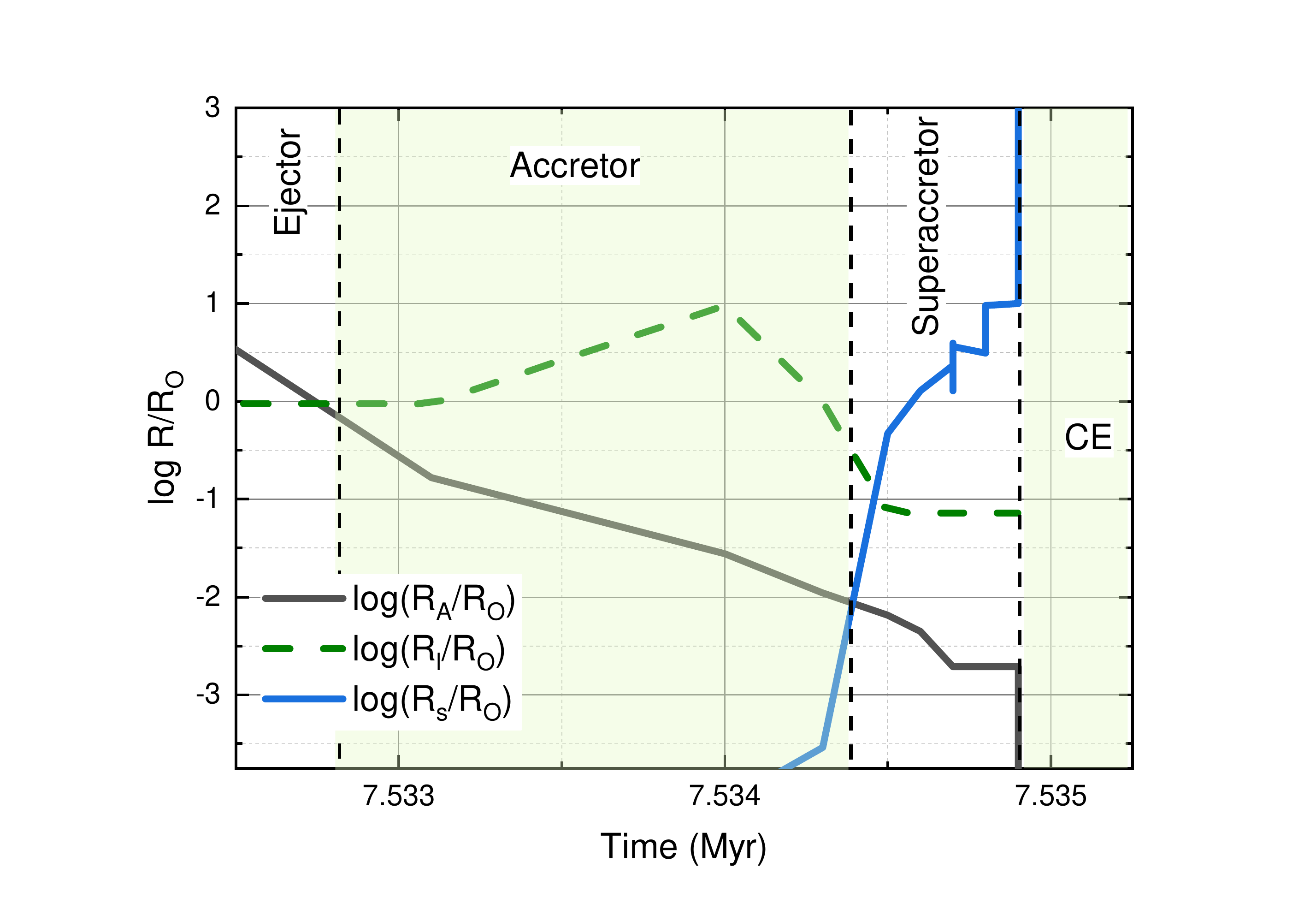}
\vfill
\includegraphics[width=0.5\textwidth]{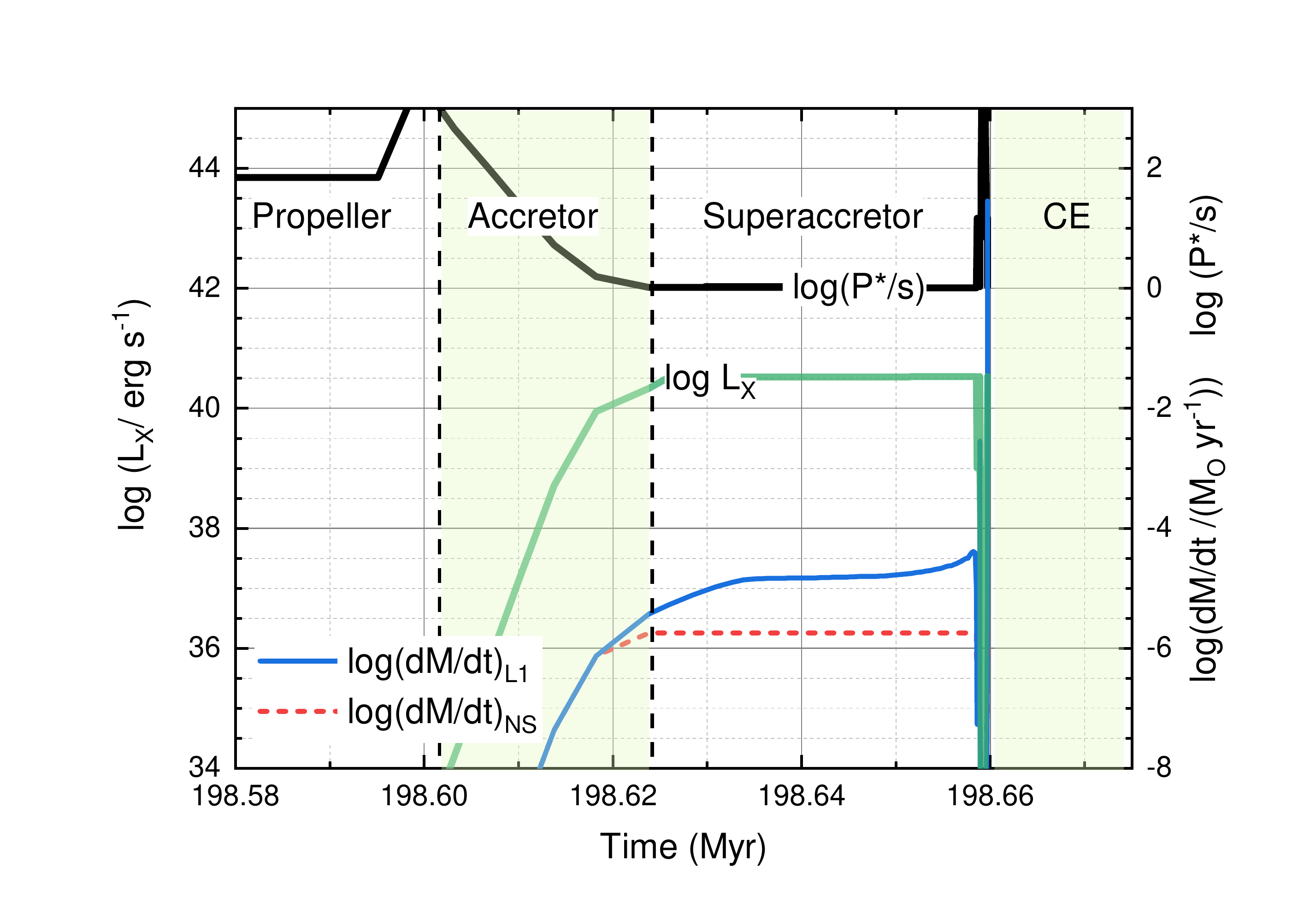}
\includegraphics[width=0.5\textwidth]{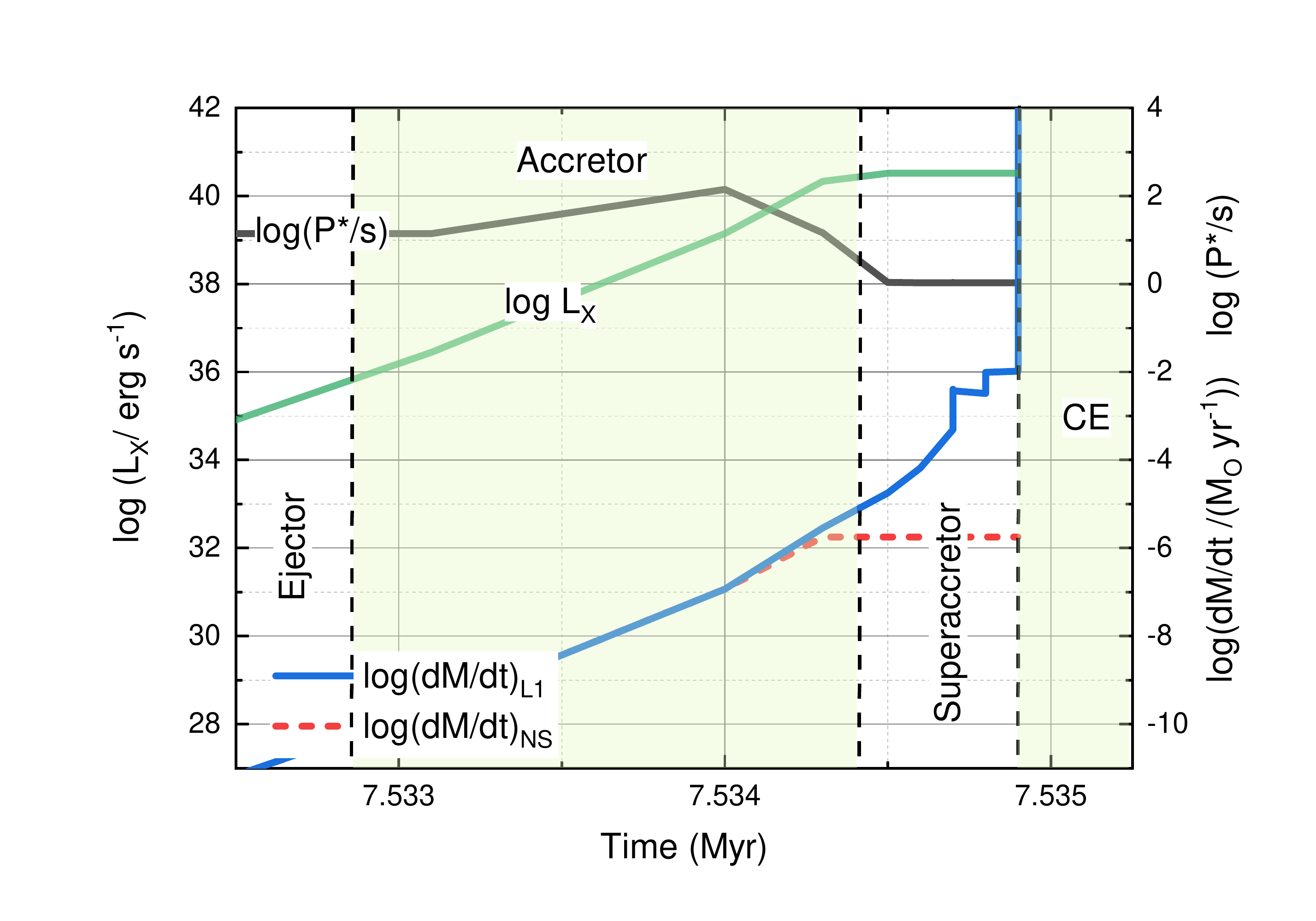}
\caption{Examples of evolutionary tracks at the stages of sub- and supercritical accretion computed by MESA. 
In the left panels, at RLOF the system ``MS-star ($M_2=3.5 M_\odot$) + NS'' has the orbital period $P_\mathrm{orb}=5$~day. In the right panels, the mass of the donor at RLOF is $M_2=21.4 M_\odot$; the orbital period is $P_\mathrm{orb}=580$~day. 
In both cases, evolution ends up by formation of the common envelope. See the text for more details. 
}
\label{fig:trc_1} 
\end{figure}

Results of computations of the duration of the stage with accretion rate exceeding 
$10^{-7}$\,\ms\,yr$^{-1}$ (which is the lower limit capable of potentially producing NULX in the case of NS with low magnetic fields)
for the grid of models
with initial donor masses (0.75 - 50)\,\ms\ and orbital semiaxes $0.9< \log (a_\mathrm{ini}/R_\odot) < 3.5$ 
are  presented in Fig.~\ref{fig:mesa_grid}
 in the coordinates $M_2 - P_\mathrm{orb,ini}$. The squares in this Figure represent the systems in which NULX stage ends up with the formation of a common envelope. 
The parameters of two tracks with parameters close to those of observed objects are shown in Fig.~\ref{fig:trc_1} and are described in detail below. 
Filled circles in the Figure show the systems which experience stable mass transfer over the entire stage of NULX. The masses of donors in such systems are below 5\,\ms, while most of 
the known sources have more massive donors. An example of a NULX with stable mass-transfer is presented in the Appendix (Fig.~\ref{fig:trc_2.5}).

In the system shown in the left panels in Fig.~\ref{fig:trc_1} the initial mass of the donor is 3.5\,\ms, the orbital period at the RLOF is 5 \,days, the NS magnetic momentum is 
$\mu_0=4\times 10^{31}$ G cm$^3$. In the upper panel, we show the relation between the mass-loss rate by the donor $(dM/dt)$ and the binary orbital period.
The colour scale reflects donor’s mass. 
The time dependence of the Alf\'{v}en radius ($R_{\rm A}$), corotation radius  ($R_{\rm co}$), and spherisation radius  ($R_{\rm s}$) 
is shown in the middle panel. 
The lower panel shows the time dependence of the NS spin period $P^\star$, 
its X-ray luminosity $L_\mathrm{X}$ given by Eqs.~(\ref{e:lx_na}) and (\ref{e:lx_nsa}) at the stages of sub- and supercritical disc accretion, respectively, the mass-loss rate by the donor and the accretion rate onto NS. The stages of subcritical accretion are shadowed. The value 
$(dM/dt)_{\rm NS}$ defines the mass rate through the magnetosphere at the supercritical stage and NS accretion luminosity  $L_\mathrm{X}=0.1(dM/dt)_{\rm NS}c^2$. The critical accretion rate $\dot M_\mathrm{cr}$ was computed using Eq.~(\ref{e:Mcr}).

Since the mass of the donor is substantially higher than the mass of accretor, $(dM/dt)_{\rm L1}$ rapidly exceeds the critical accretion rate onto NS $\dot M_\mathrm{cr}$. The middle and bottom panels in Fig.~\ref{fig:trc_1} clearly
show that formally the system can manifest itself as NULX 
($L_{\mathrm X} {>} 10^{39}$\,\lum) already at the short subcritical accretion stage.   Mass loss by the donor and NULX stage terminate by formation of the common envelope. In this case, the stage of NULX lasts for $\simeq 40000$~yr. 

The right panels in Fig.~\ref{fig:trc_1} show an example of the track of visual star that has at RLOF the mass  $M_2=21.4 M_\odot$ 
 and the orbital period  $P_\mathrm{orb}=564$~day. The NS magnetic field is the same as in the example to the left. In this system, NULX stage with 
$L_{\rm X} > 10^{39}$\,\lum\ is short (less than 1000~yr, see the middle and lower panels) and also ends up by the common envelope formation. 
Note, unlike the system to the left, the NS spin period $P^*$ at the accretion stage
(the upper line in the right lower panel) first increases to the equilibrium value 
$P_{eq}$ when the age of the system is 7.534~Myr, and afterwards the NS spins up according to $P^*=P_{eq}\sim \dot M^{-6/7}$. 
In this case, this is related to the transition of NS from the ejector to the accretor stage, avoiding the propeller stage.

\begin{figure} 
\includegraphics[width=0.49\columnwidth]{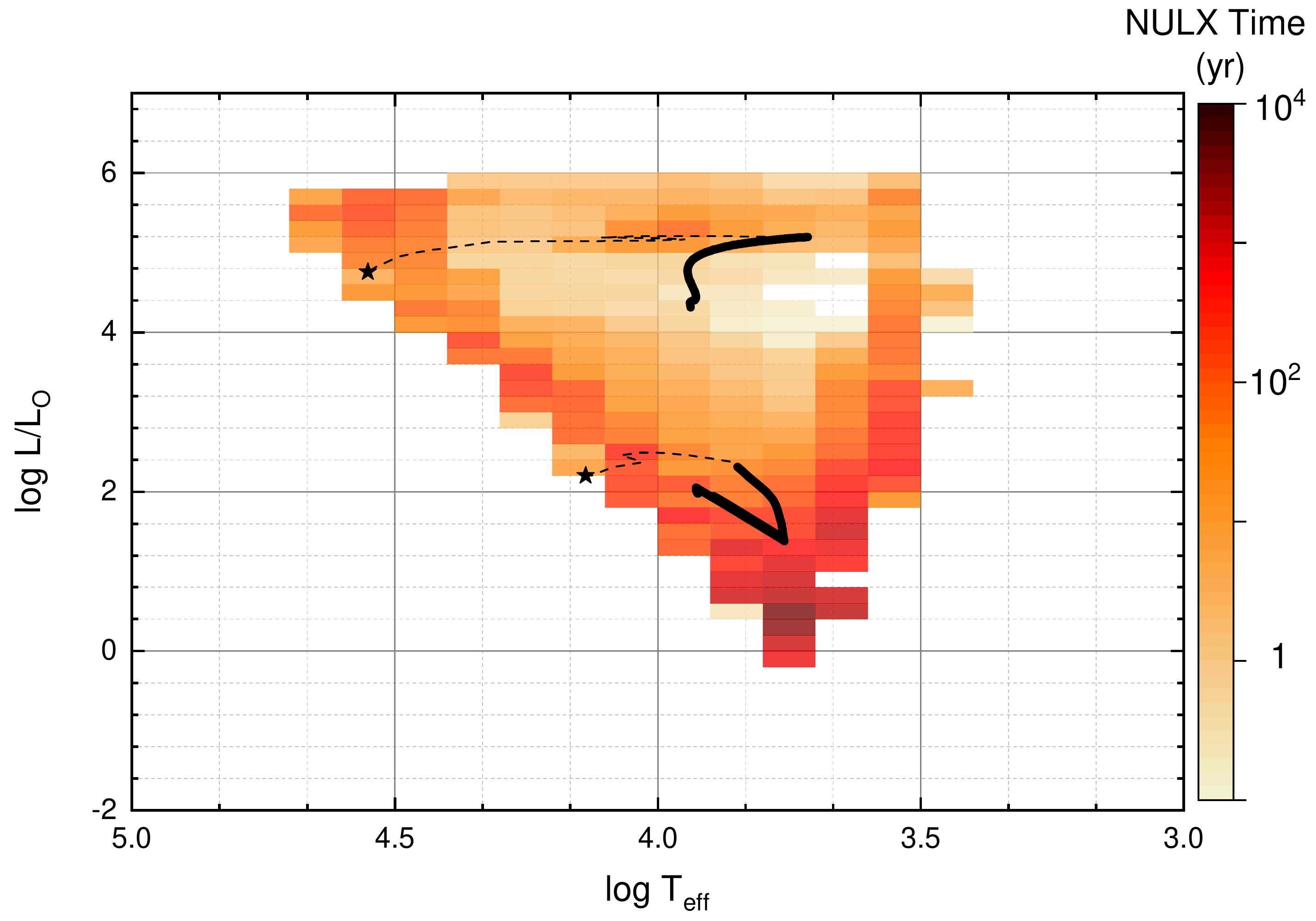}
\includegraphics[width=0.49\columnwidth]{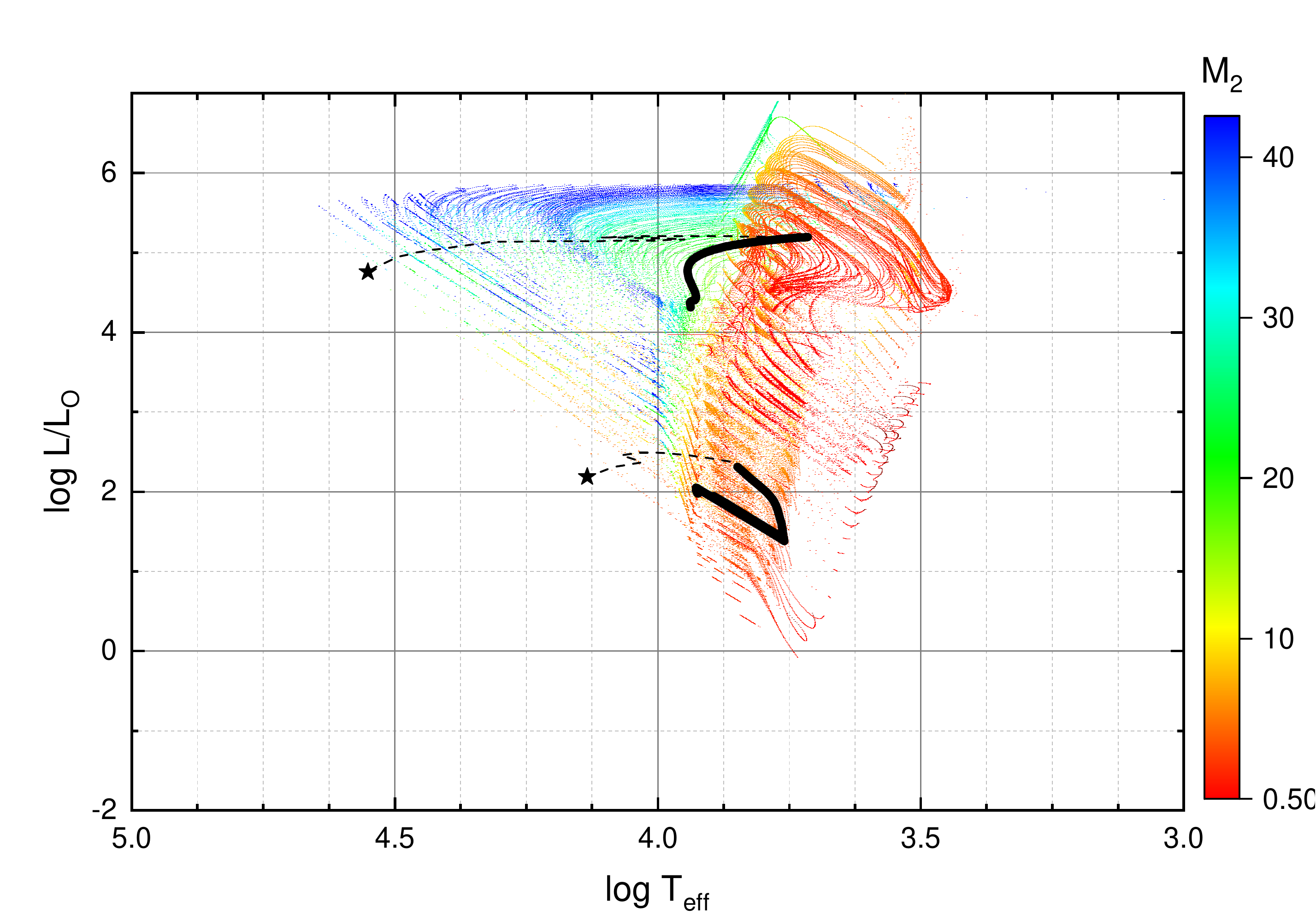}
\caption{Position of the optical components  ($M_2$) in the Hertzsprung-Russel diagram during NULX stage. In the left panel, the NULX stage duration is shown in colour. In the right panel, the colour of points reflects the mass of the visual component. Black lines mark the tracks of the stars with mass 3.5 
and 21.4\,\ms\ presented in Fig.~\ref{fig:trc_1} at the RLOF stage. Solid segments of the lines show the stages when the systems are NULX.
}
\label{fig:HR}
\end{figure}
Figure~\ref{fig:HR} shows location of the optical components of binaries that are NULX in the  Hertzsprung-Russel diagram. It is evident that the overwhelming majority of them, despite long time duration of the stage, cannot be discovered by modern instruments due to low luminosity. Known visual components of NULX have masses   
$\apgt 5$\,\ms (see Table~1). The luminosity of 5~\ms\ stars is close to 500\,\ls. Thus, the stars in the most populated part of the HR diagram are cut off. 
At the same time, Fig.~\ref{fig:HR} shows that a significant fraction of donors in NULX should be red (super)giants. For instance, the optical component of  NGC30~ULX-1 is a red giant  (Heida et al. 2019).

\subsection*{Model diagram $M_2 - P_\mathrm{orb}$}

\begin{figure} 
\includegraphics[width=\columnwidth]{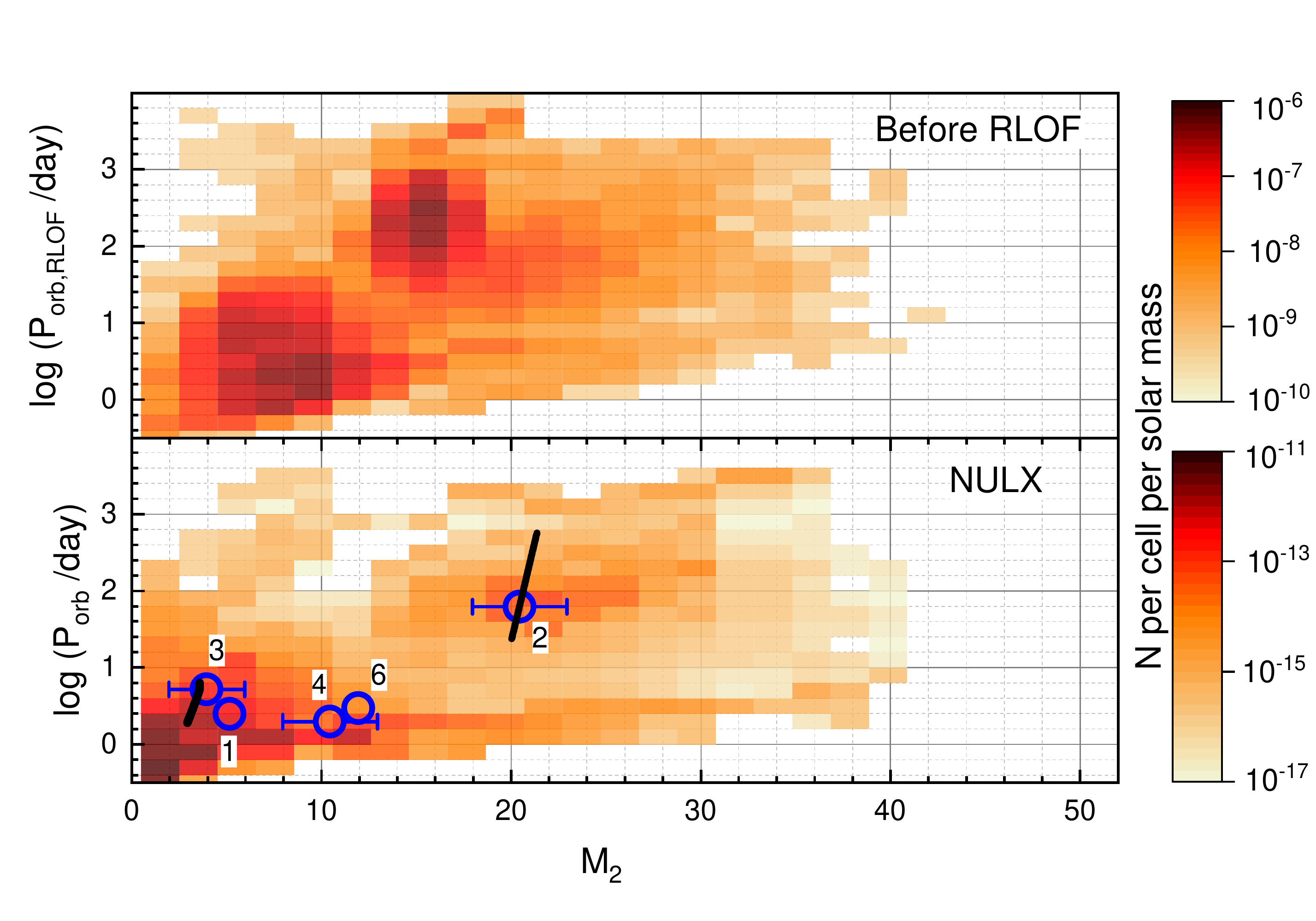}
\caption{Model diagram ``mass of the visual star $M_2$ -- orbital period 
$P_\mathrm{orb}$''.
Upper panel: distribution of the number of CBS with NS at RLOF by donors calculated by BSE from the evolution of  $10^7$ initial binaries with distributions of initial parameters described in the text. 
Colour scale shows the number of systems normalised to solar mass.
Lower panel: distribution of NULX in the model galaxy normalised to the solar mass, obtained by convolution of the frequency of formation of their precursors (code BSE, upper panel) with the time length of NULX phase (code MESA, see Fig.~\ref{fig:mesa_grid}) and star formation history (Eq.~\ref{e:SFR})  14~Gyr after the beginning of star formation in the model galaxy.  
 Magnetic fields of NS are defined by Eq.~(\ref{e:mu}) with the mean value 
$\mu_0 = 10^{31.6}$~G cm$^3$.
Circles -- quasi-stationary sources (Table 1). Black lines -- the segments of model tracks shown in Fig.~\ref{fig:trc_1} at the NULX stage.  }
 \label{fig:bse_map}
\end{figure}

Distribution of the number of systems that are potential precursors of NULX at the instant of RLOF in the ``visual component mass -- orbital period'' 
($M_2 -  P_\mathrm{orb}$) diagram, as computed by the code BSE for $10^7$ initial binaries is shown in the upper panel of Fig.~\ref{fig:bse_map}. The shortage of systems with donor masses  (12 -- 14)\,\ms\ and $\log(P/\hbox{day})\apgt 1.6$ is related to the transition from MS donors to Hertzsprung gap and red-giant donors. The similar deficit, but for slightly lower masses, was noted earlier by Fragos et al. (2015). 

The convolution of the formation probability of a close binary with NS and optical component with mass $M_2$ at the verge of RLOF, computed per one solar mass by BSE  (upper panel of Fig.~\ref{fig:bse_map}) with the time duration of NULX calculated for a grid of MESA models and the star formation rate (Eq.~(\ref{e:SFR})), yields the expected number of NULX per one solar mass in the model galaxy (the lower panel of Fig.~\ref{fig:bse_map}). 
In the lower panel of the Figure, two groups of sources can be distinguished – with masses of optical components of a few solar masses and orbital periods up to $\simeq 10$~day and sources with masses of the order of 20\,\ms\ and orbital periods from several tens to about 100 days. Examples of evolutionary tracks populating these regions of the diagram were described above and are shown in Fig.~\ref{fig:trc_1}. NULX stages for these sources are plotted in the Figure by solid lines. 
It is well seen that positions of the sources NGC7793~P13 and NGC5907~ULX-13  (Table 1) are close to model tracks shown in the left and right panels of Fig.~   \ref{fig:trc_1}, respectively. 
\begin{figure}        
\includegraphics[width=0.49\columnwidth]{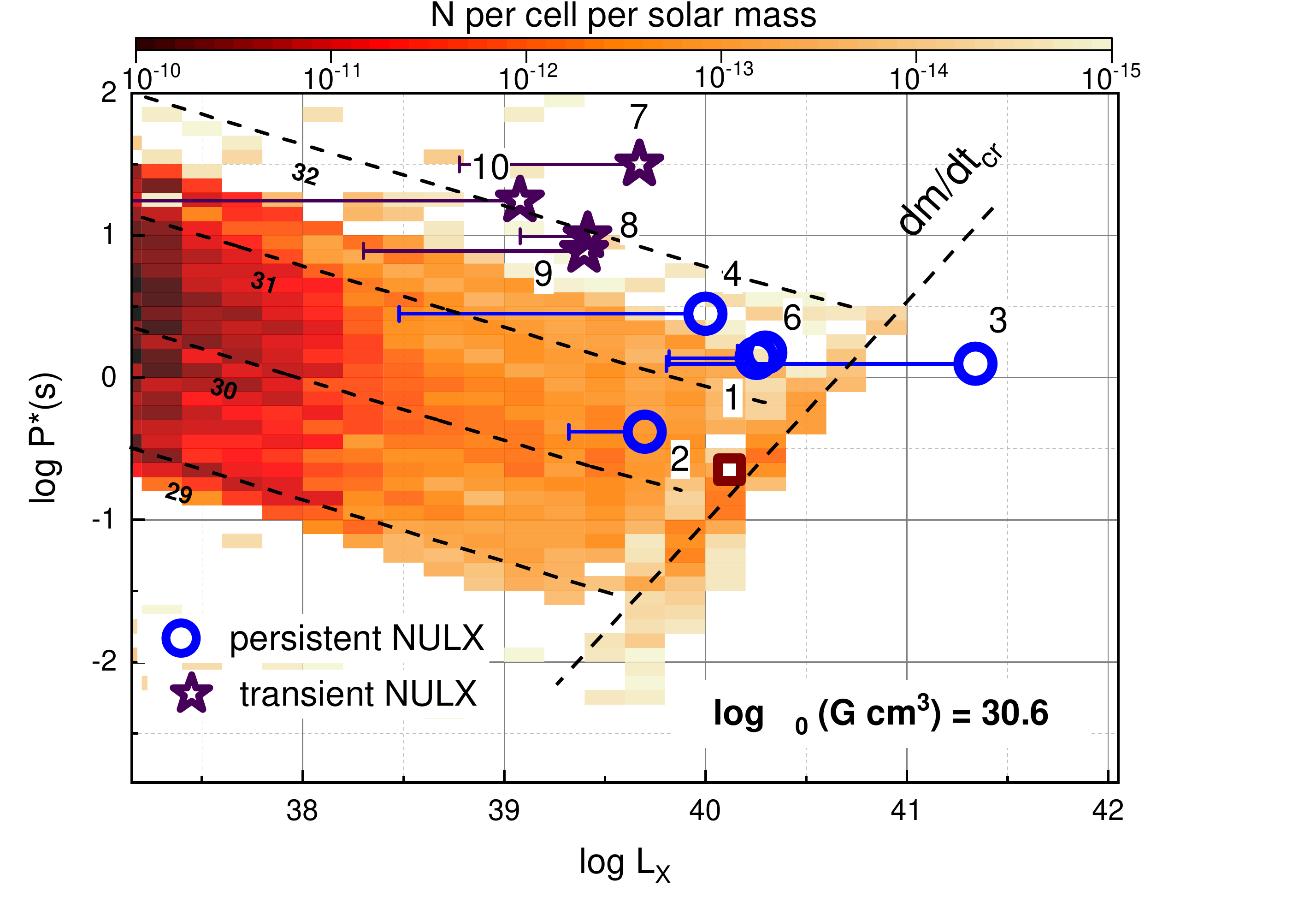}
\includegraphics[width=0.49\columnwidth]{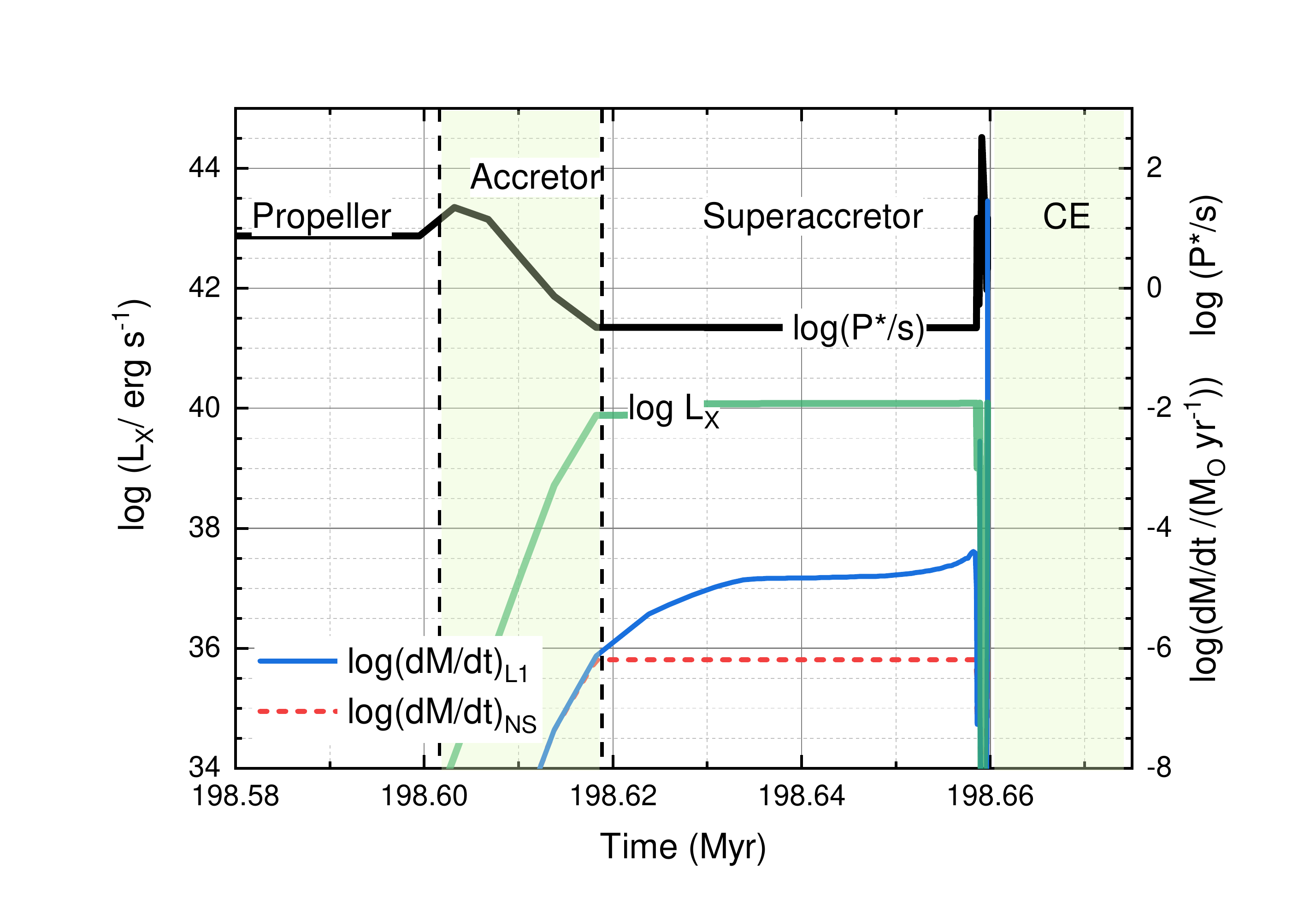}
\vfill
\includegraphics[width=0.49\columnwidth]{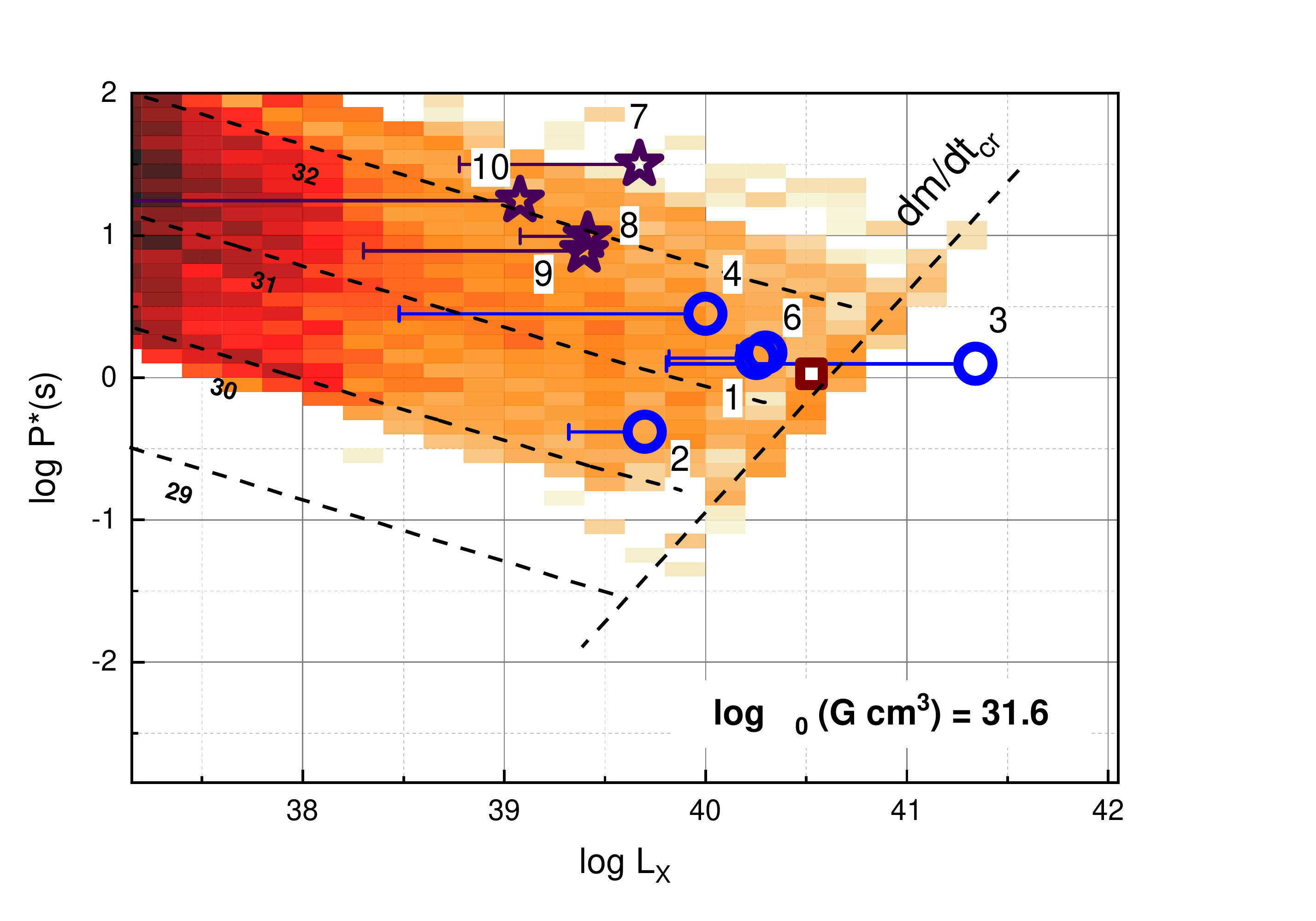}
\includegraphics[width=0.5\textwidth]{trc_m35_g3_3.pdf}
\vfill
\includegraphics[width=0.49\columnwidth]{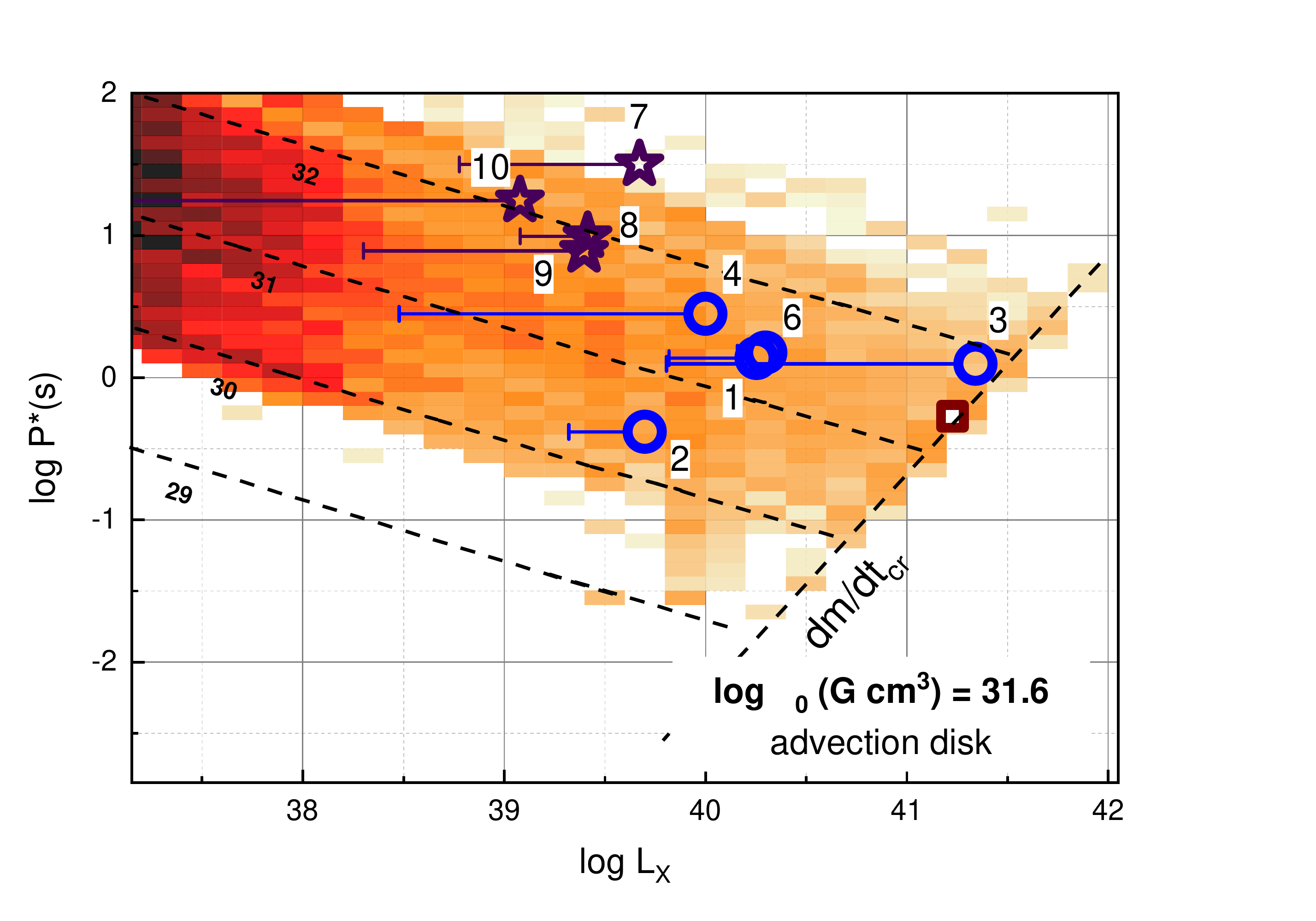}
\includegraphics[width=0.49\columnwidth]{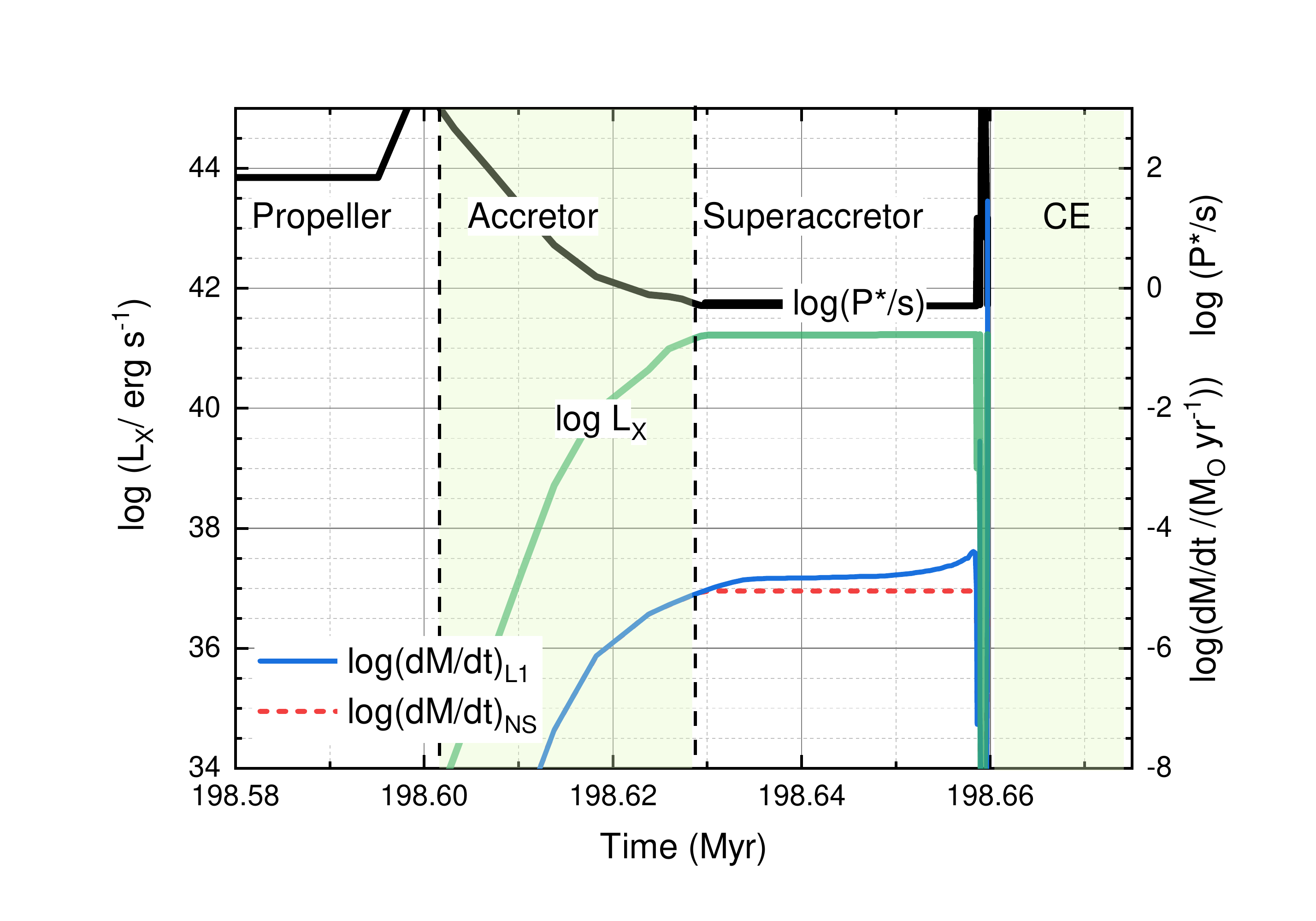}
\vfill
\caption{ The distribution of sub- and supercritical accreting NS over spin periods and X-ray luminosity. 
Upper left panel: the model of self-regulating supercritical accretion disc. Magnetic field strength is defined by Eq.~(\ref{e:mu}) with mean $\mu_0 = 10^{30.6}$~G~cm$^3$. 
Left middle panel -- the same as in the upper panel but for 
$\mu_0 = 10^{31.6}$~G~cm$^3$.
Lower left panel: the same for the supercritical advective disc model and 
$\mu_0 = 10^{31.6}$~G~cm$^3$.
Observed NULX are shown by open circles and asterisks and annotated by the numbers following Table~1. Open square – position of the model supercritical source with the initial mass of the donor and orbital period like in the left panel of Fig.~\ref{fig:trc_1}, 
but in different models. 
Right panels: characteristics of the model NULX (the square) for different magnetic fields in the upper two panels and advection disc model in the bottom panel. 
}
\label{fig:na}
\end{figure}

\subsection*{$L_\mathrm{X}-P^*$ diagram for accreting NS}

Model diagrams ``X-ray luminosity $L_\mathrm{X}$ -- spin period of NS $P^*=P_\mathrm{eq}$'' 
for NS at sub- and supercritical accretion stages are shown in 
Fig.~\ref{fig:na} (panels to the left). Colour scale shows the number of systems per solar mass. We present results for three models: disc accretion with a log-normal distribution of magnetic momenta with mean $\log\mu_0=30.6$~G~cm$^3$ corresponding to the surface magnetic field of NS with radius 12~km 
$B_0\approx 5\times 10^{12}$ G (upper panel), the same model but with an order of magnitude higher magnetic field ($\log\mu_0=31.6$~G~cm$^3$, $B_0\approx 5\times 10^{13}$~G, middle panel), and the model of advective disc around NS with
mean magnetic moment $\log\mu_0=31.6$~G~cm$^3$ (bottom panel). 
Dashed lines marked by the values of NS magnetic moments correspond to the dependence of the NS equilibrium spin $P_\mathrm{eq}$ on the accretion rate 
(X-ray luminosity) for subcritical disc accretion (Eq.~(\ref{e:peqa})). We assume that for the slow evolution of accretion rate the NS spin adjusts to the equilibrium value   
($P^*=P_\mathrm{eq}(L_\mathrm{X})$) and, consequently, the spin evolution proceeds along these lines. 
Dashed straight line $(dM/dt)_{cr}$ corresponds to the critical accretion rate 
$\dot M_\mathrm{cr}$
(NS X-ray luminosity $L_\mathrm{X}$ after Eq.~(\ref{e:lx_nsa})). Thus, the sources along this line and to the right of it are at the supercritical accretion stage. 
Circles and asterisks marked by the numbers like in Table~1 show positions of observed quasi-stationary (persistent) and transient NULX. 
The red square shows the position of supercritical NULX with an initial mass of the donor and orbital period as in the left column in Fig.~\ref{fig:trc_1}.  
Panels to the right show results of MESA computations for this model source – NS spin period, X-ray luminosity, mass-transfer rate via $\mathrm{L_1}$ and accretion rate onto NS -- for the NS magnetic momentum $\mu=10^{30.6}$ and $10^{31.6}$~G~cm$^3$ and models of self-regulating supercritical accretion (two upper panels) and advection disc (lower panel).
For the most time of RLOF by the donor, such sources are at the stage of supercritical accretion. 
\label{s:results}

\subsection*{Distribution of the parameters of model NULX}

\begin{figure*} 
	\includegraphics[width=0.9\textwidth]{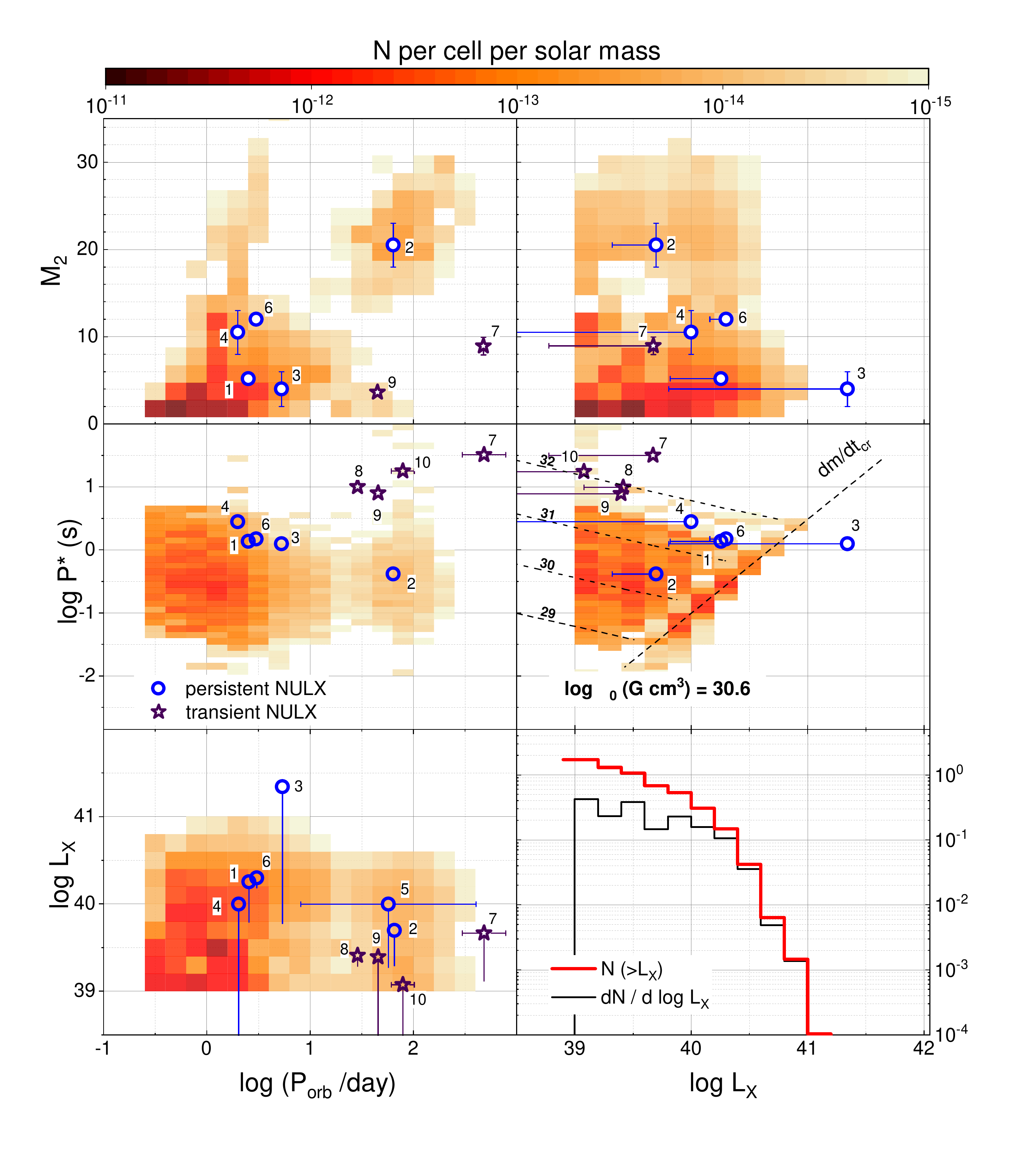}
	\caption{The right middle panel: model distributions of NULX  at the stages of sub- 
and supercritical accretion, to the left and to the right of the line $(dM/dt)_{cr}$,
respectively. Panels from top to bottom and from left to right show relations 
$M_2-\log P_\mathrm{orb}$, $\log P^*-\log P_\mathrm{orb}$, $\log L_\mathrm{X}-\log P_\mathrm{orb}$, $\log M_2-\log L_\mathrm{X}$, $\log P^*-\log L_\mathrm{X}$. The right lower panel shows cumulative ($N>L_\mathrm{X}$) and differential 
($dN/d\log L_\mathrm{X}$) X-ray luminosity functions of NULX  (per solar mass in the model galaxy, scale to the right). The plots are normalised to the current (at the age of 14\,Gyr) 
SFR=5\,\ms\,yr$^{-1}$.
Colour scale at the top of the Figure -- the number of sources in the cell of the grid per solar mass. The mean NS magnetic momentum is 
$\log\mu_0 = 30.6$\,G\,cm$^3$.}
	\label{fig:b12}
\end{figure*}

\begin{figure*}   
	\includegraphics[width=0.9\textwidth]{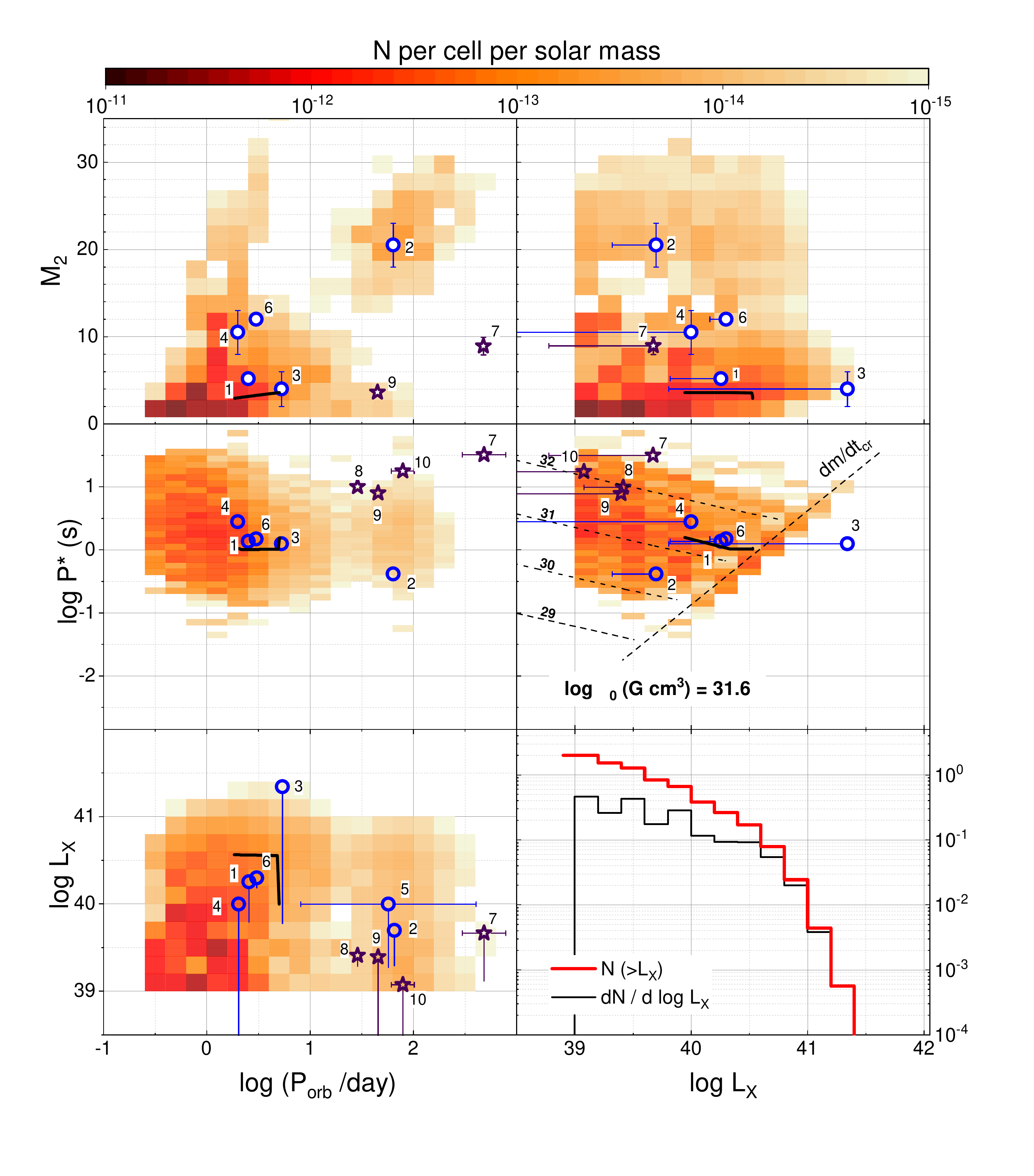}
	\caption{Same as in Fig.~\ref{fig:b12} for the NS mean magnetic momentum 
	$\log\mu_0 = 31.6$\,G\,cm$^3$. 
	 Black lines show the model track of the donor in CBS with initial mass $M_2$ and orbital period $P_\mathrm{orb}$ similar to that in the left column in Fig.~\ref{fig:trc_1}.
	 }
	\label{fig:b13}
\end{figure*}

\begin{figure*} 
	\includegraphics[width=0.9\textwidth]{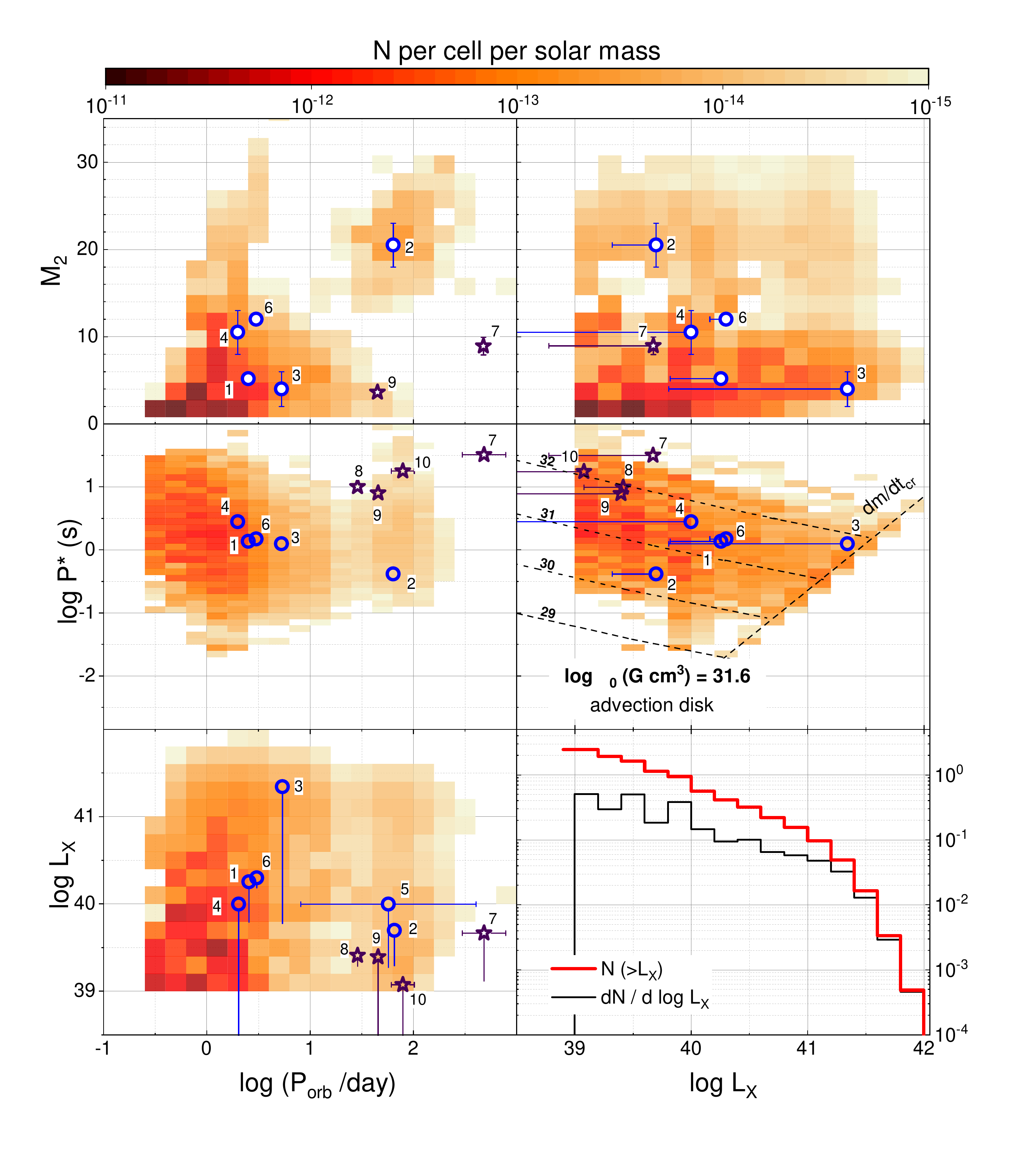}
	\caption{Same as in Fig.~\ref{fig:b12} for advective disc around magnetised NS (Chashkina et al. 2019) and the NS mean magnetic momentum 
	$\log\mu_0 = 31.6$\,G\,cm$^3$. 
	}
	\label{fig:b13a}
\end{figure*}

In our models, NULX represent a subset of NS at the stage of disc accretion in CBS with luminosity $L_\mathrm{X}>10^{39}$\,\lum. 
Their distributions over observed parameters -- masses of visual donor-stars
$M_2$, NS spin periods $P^*$, X-ray luminosity  $L_\mathrm{X}$, and orbital periods of CBS $P_\mathrm{orb}$
are plotted in two-dimensional colour diagrams in 
Figs.~\ref{fig:b12}-\ref{fig:b13a} for the same three models as in the Fig.~\ref{fig:na}. Colour scale shows the number of systems per solar mass in the model galaxy with star formation rate (\ref{e:SFR}) at the age of 14~Gyr. 
Like in Fig.~\ref{fig:na}, circles and asterisks show positions of observed quasi-stationary and transient NULX. In the right lower panel, X-ray luminosity functions  of NULX are plotted (differential  $dN/d\log L_\mathrm{X}$ and
cumulative $N(>L_\mathrm{X})$). Analysis of X-ray observations suggests (Mineo et al. 2012, Sazonov and Khabibullin 2017) that the number of high-mass X-ray binaries in the galaxies is proportional to the star formation rate (SFR). For this reason, in the scale to the right, we show normalisation of
our results for average SFR in the model galaxy  
$\langle \mathrm{SFR}\rangle=5 M_\odot$ yr$^{-1}$, as it follows from Eq.~(\ref{e:SFR}), at the age  13-14~Gyr. Here, the mass  of stellar disc  $10^{11} M_\odot$ corresponds to SFR$\approx 7\,  M_\odot$ yr$^{-1}$. 
 
It is seen that even for conservative assumptions about the character of disc accretion and standard mean magnetic field of NS (Fig.~\ref{fig:b12}), the model reproduces the observed location of quasi-stationary NULX (circles).
Note that the majority of NULX can be explained by disc accretion onto NS with magnetic field in the range $10^{12}<B<10^{14}$~G (see the plot  $L_{\rm X}$-$P^*$ in the middle right panel).                                                          
Only one source (NGC5907~ULX1) falls into the region of supercritical accretion onto NS with magnetic field $\sim 10^{13}$~G. 
Note also that transient NULX (asterisks) with the super-Eddington luminosity during outbursts and with a low quiescent X-ray luminosity fall into the most densely populated area of Fig.~\ref{fig:na}.

Increase in the mean magnetic field  of NS  to $\sim 10^{13}$~G
 (Fig.~\ref{fig:b13}) increases the NS equilibrium periods (Eqs.~(\ref{e:peqa}), (\ref{e:peqsa})), as well as the limiting X-ray luminosity (\ref{e:lx_nsa}) and extends the  
X-ray luminosity function for the brightest sources beyond $10^{41}$\,\lum. 
In the model of supercritical advective discs with high accretion rate (\ref{e:mcrad}), the limiting possible accretion luminosity of NULX becomes higher than $\sim 10^{41}$\,\lum. However, X-ray luminosity of such sources may appear lower
(see the discussion of limiting luminosities in the model accretion columns in 
Mushtukov et al. (2017)).  Note, however, that the assumed structure of accretion discs only weakly affects the location of model sources in the ``mass of the visual star -- orbital period’’ diagram.


\section*{Discussion and Conclusion}
\label{s:conclusion}

After the discovery of the first pulsing ULX  (Bachetti et al. 2014) it became clear that a significant fraction of the population of ULX may harbour accreting NS 
(see, for instance, Koliopanos (2017), Walton (2018)).  Population synthesis studies confirm this inference, but they have been done without an account of the magnetic fields of NS. 
Here we show that accretion onto magnetised NS naturally reproduces observed features of NULX without additional hypotheses on the beaming of X-ray radiation from accreting NS (cf.  Wiktorowicz et al. 2017).

We have used for computations a hybrid method of population synthesis. Potential precursors of the objects of interest (NULX) were found by calculating the evolution of a large number $(10^7)$ of close binaries with given initial parameter distributions using rapid analytical population synthesis code BSE (Hurley et al. 2002).  
Next, the mass transfer rates were computed using a grid of MESA models (Paxton et al. 2011) with an accurate account of stellar evolution and physics of mass-transfer in CBS. The novelty of our computations is a detailed treatment of X-ray luminosity during disc accretion onto magnetised NS. For high accretion rates, super-Eddington luminosities 
$L_{X}>10^{38}$\,\lum\ are attained already at the stage of subcritical accretion in the disc, while the supercritical phase begins only when the local Eddington luminosity is achieved at the inner rim of the disc close to the NS  magnetosphere ($R_A> R_{NS}$).   
It is not surprising that in the standard model of the CBS evolution,  for NS with magnetic fields of $10^{12}-10^{14}$~G X-ray luminosities up to $10^{40}-10^{41}$\,\lum\ (see Fig.~\ref{fig:na}) are realised in a natural way. 

An essential factor which defines the characteristics of the accretion stage onto NS  (its time duration, accretion rate, the possibility of formation of a common envelope) 
is the non-conservativeness parameter $\gamma_\mathrm{mt}$ (dimensionless angular momentum brought away by the matter leaving the system via the outer Lagrange point  
$\mathrm{L_2}$). We assumed that in this way a modest fraction of the matter lost by the donor leaves the system, 
$\delta_\mathrm{mt}=0.1$. 
However, if this matter has a minimum possible value of $\gamma_\mathrm{mt}$=1.15, the orbital periods of observed NULX are not reproduced. For the value  $\gamma_\mathrm{mt}=3$ assumed by us, which is almost twice as low as that derived from observations of SS433 (Cherepashchuk et al. 2019), the agreement with observations becomes satisfactory,    
see the diagrams  $P^*-L_\mathrm{X}$, $P^*-P_\mathrm{orb}$, $P_\mathrm{orb}-L_\mathrm{X}$, $M_2-P_\mathrm{orb}$ and $M_2-L_\mathrm{X}$ (Figs.~\ref{fig:b12}-\ref{fig:b13a}) for different accretion disc models and NS mean magnetic fields. 

For completeness, in the  $P^*-L_{\rm X}$ diagram we marked the position of transient X-ray sources. Their evolution differs from that of quasi-stationary sources: during the outbursts, the transients rapidly move in the horizontal direction in this diagram, their spin periods being the NS equilibrium values in the quiescence state where they spend most of the time. Depending on the amplitude of the outburst, a transient NULX may, for instance, move out from the densely populated region of persistent low-luminosity NS accreting matter from the discs or even from the quasi-spherical stellar winds, see computations in Postnov et al. (2019).

In addition to the standard accretion discs, we considered the model of advection disc around magnetised NS (Lipunova 1999, Chashkina et al. 2019) which allows even larger critical accretion rate onto NS (Eq.~(\ref{e:mcrad}) and lower panel in Fig.~\ref{fig:na}). However, the possibility of X-ray luminosity from accretion columns exceeding $10^{41}$\,\lum\ remains debatable from theoretical point of view  (Mushtukov et al. 2017). Future observations of NULX with super-high luminosities would be essential to test accretion models onto magnetised NS.

Apart from the explanation of the observed X-ray luminosities and NS spin periods, the CBS models computed by us also reproduce known
orbital periods and estimated masses of optical components of NULX   (see Figs.~\ref{fig:b12}--\ref{fig:b13a}).
The luminosity function presented in the right bottom panels of these Figures suggests that in a model galaxy with the assumed star formation history    
 (Eq.~\ref{e:SFR}) and the mass of the thin disc of the order of that in the Milky Way or, equivalently, with the current star formation rate  
3-5 $M_\odot$ yr$^{-1}$, several NULX with luminosity up to 
$10^{40}$\,\lum\ can be found.
Note, considering the population of NULX in the model spiral galaxy, distributions with a 
high mean NS magnetic field ($\sim 5\times 10^{13}$ G) are more preferable than follows from the analysis of the population of radio pulsars 
(Faucher-Gigu{\`e}re and  Kaspi 2006), see Figs.~\ref{fig:b13} and \ref{fig:b13a}.
Though, to make any conclusions about magnetic fields of NS based on population synthesis studies would be, evidently, too far a stretch. 

We conclude that the standard evolution of close binaries with NS with a detailed treatment of mass transfer by MESA enables quantitative explanation of the observed NULX in spiral galaxies as the objects with magnetised accreting NS and does not require strong X-ray beaming. 
From the physical point of view, the most interesting can be high X-ray luminosity sources with supercritical accretion (currently, only one such candidate object is known,  NGC5907~ULX-1). 
There should be powerful outflows from supercritical discs forming envelopes and nebulae around the systems.  Discovery of
optically thick outflows from Swift J0243 (Tao et al. 2019) and expanding nebula around NGC~5907 (Belfiore et al. 2020) may be an evidence  of formation of  supercritical discs around NS.\\ 
\textit{Note added in proof}. A.D. Chandra et al. (\mnras\ {\bf 495}, 2664 (2020)) reported discovery of a pulsing transient source RX~J0209.6-7427 with $L_{\rm X}=1.6\times 10^{39}$
\lum. 

\vspace{2mm}
The authors acknowledge G.V. Lipunova for fruitful discussions and the referees for comments that helped to improve the presentation. The study was supported by the RFBR grant  19-12-00229. A.G. Kuranov and K.A. Postnov were supported by the Moscow University scientific school ``Physics of the stars, relativistic objects and galaxies''. L.R. Yungelson acknowledges partial support by the RFBR grant 19-07-01198.

\section*{Appendix}
\label{s:app}
\subsection*{The effect of the parameters of non-conservative mass-transfer}

\begin{figure}[ht!] 
\includegraphics[width=0.5\textwidth]{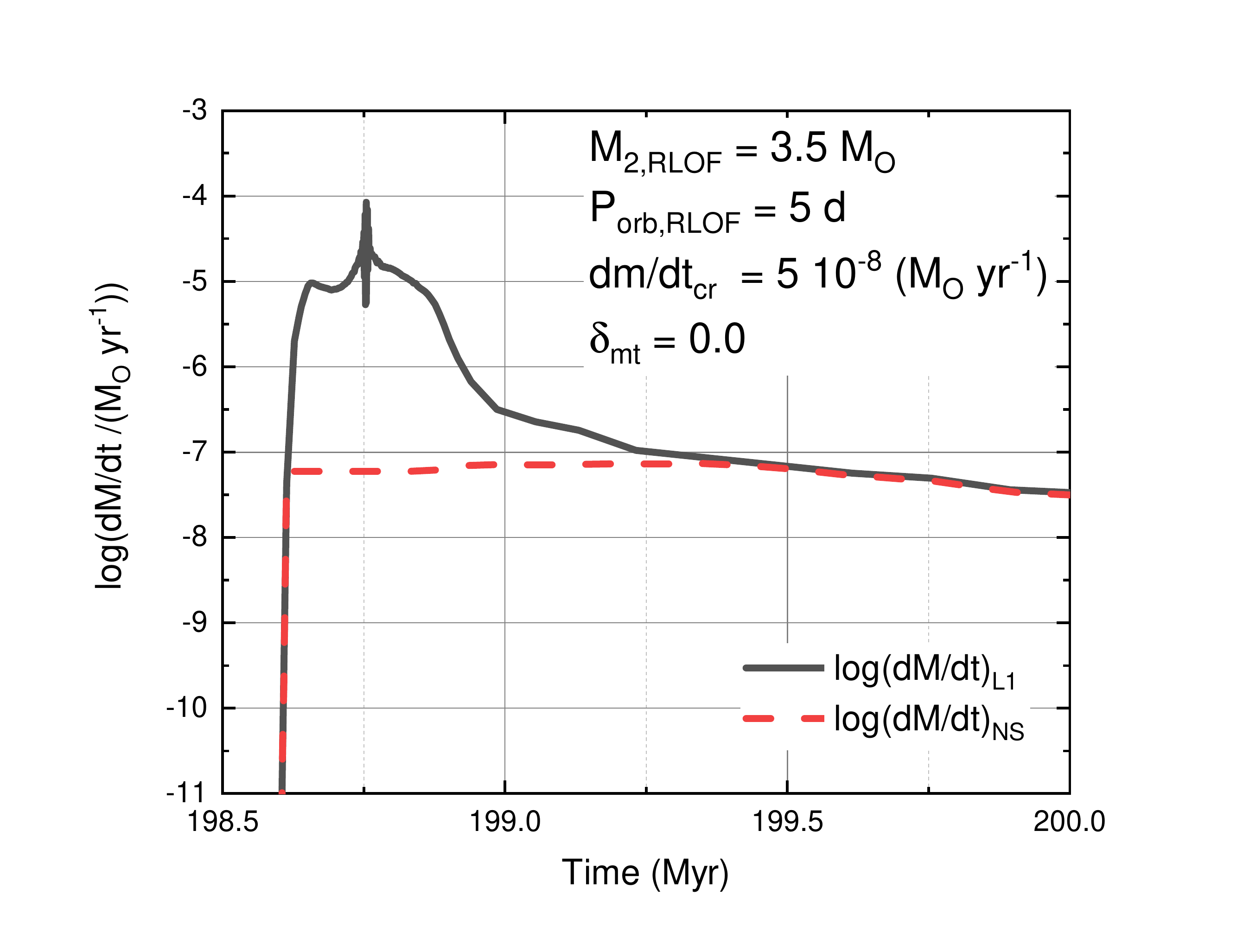}
\includegraphics[width=0.5\textwidth]{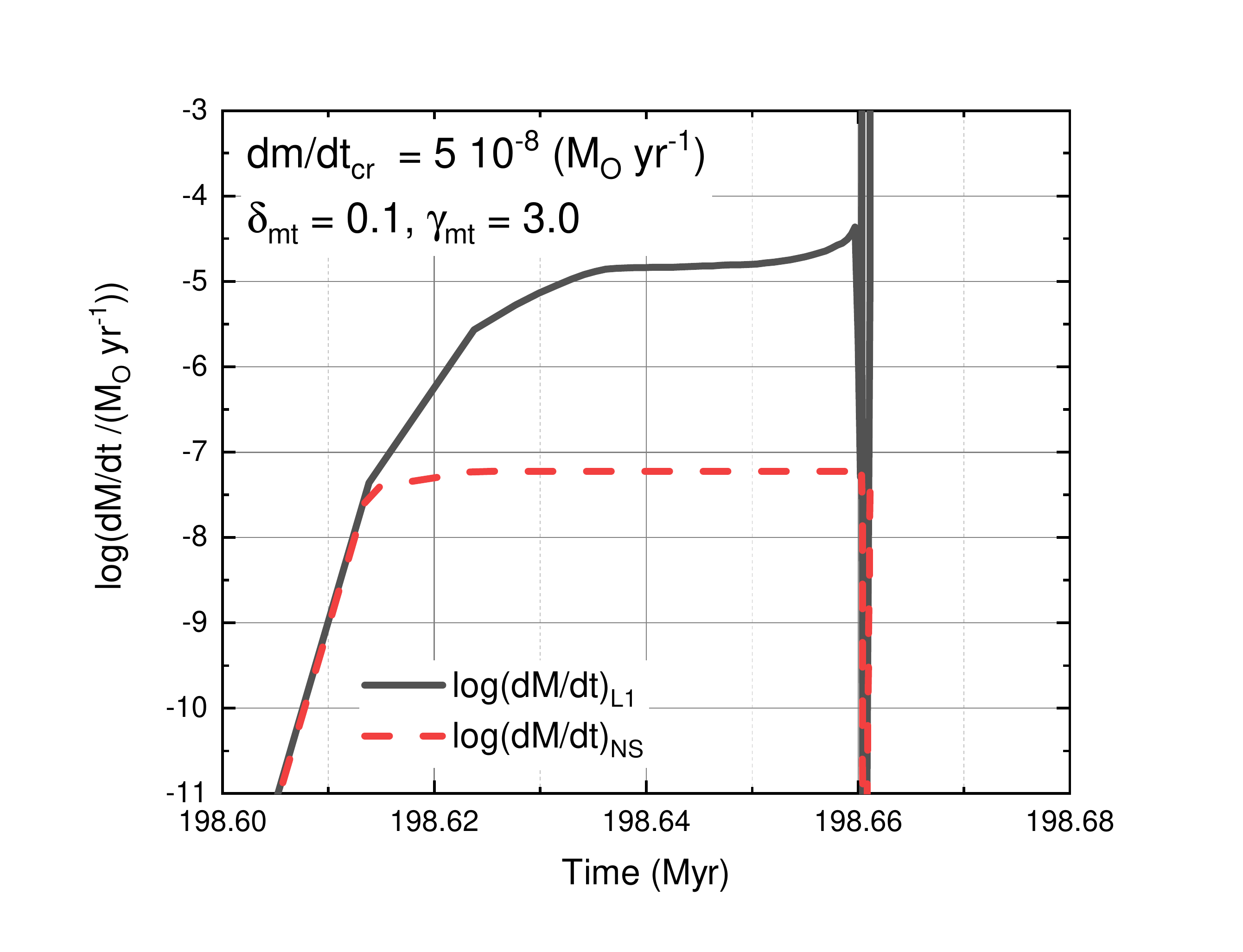}
\vfill
\includegraphics[width=0.5\textwidth]{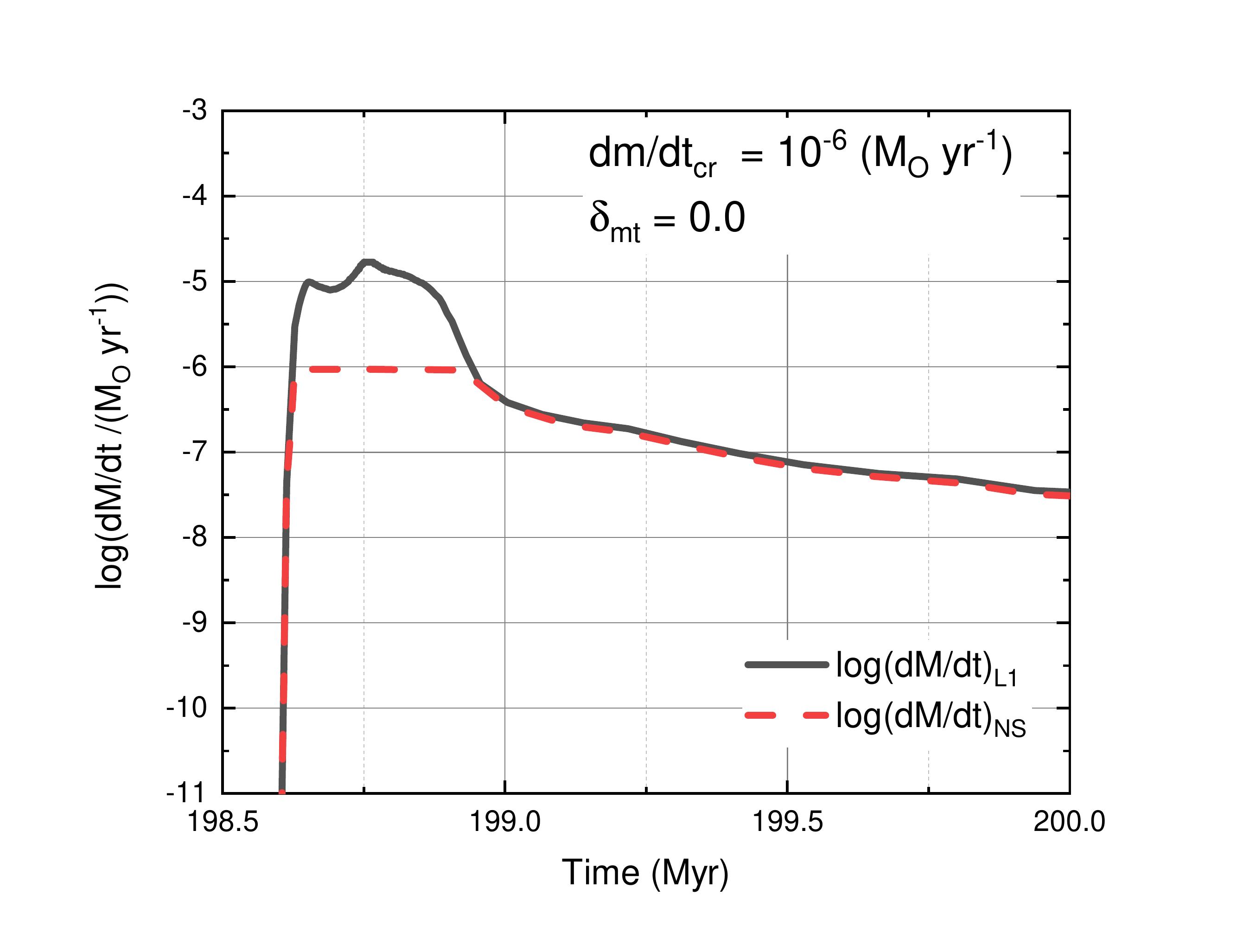}
\includegraphics[width=0.5\textwidth]{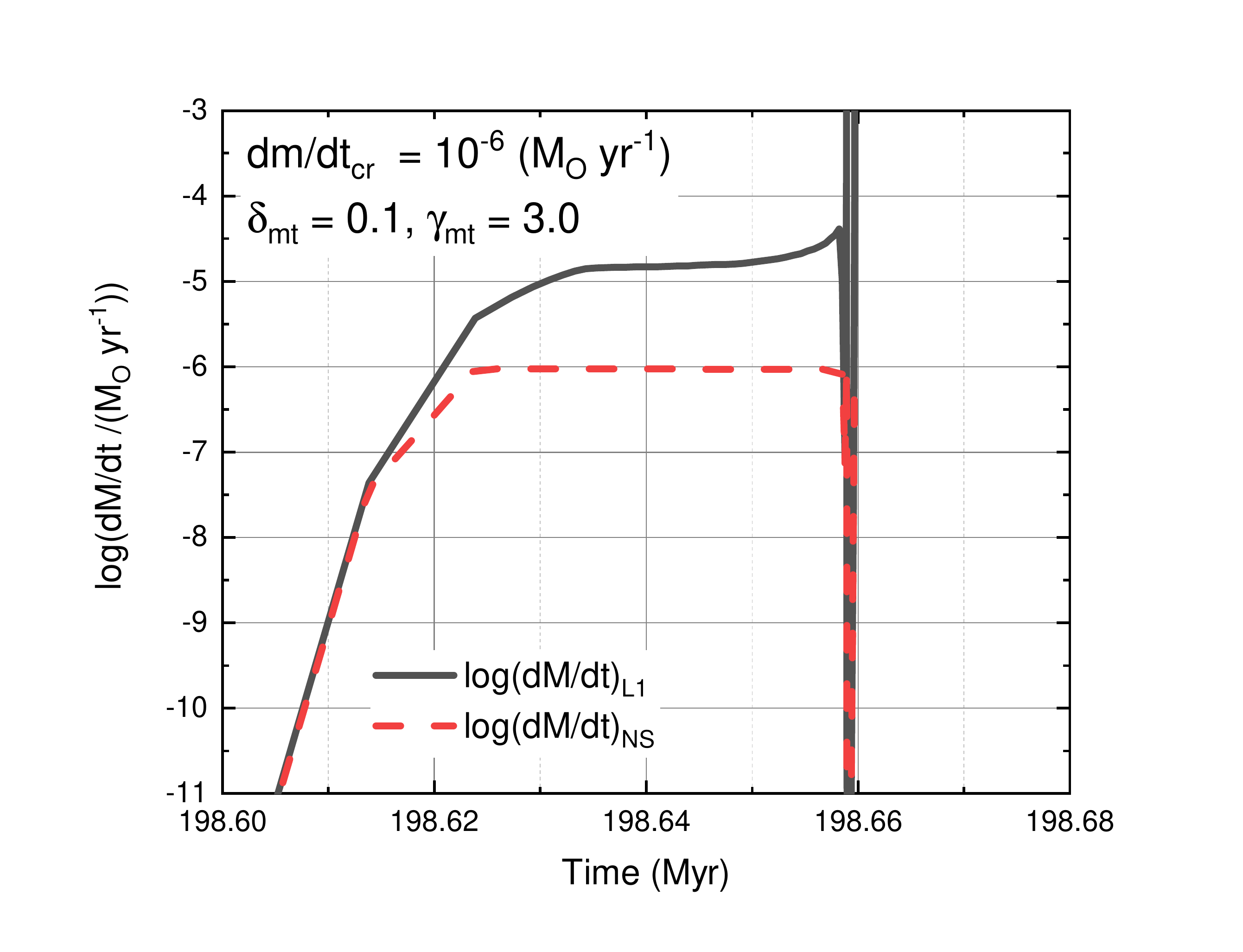}
\caption{Examples of the evolutionary tracks computed by MESA under different assumptions on the parameters of the non-conservative evolution. See the text for details. 
}
\label{fig:trc_g}
\end{figure}

Figure \ref{fig:trc_g} shows results of computations of evolution by MESA 
for the same system as in Fig.~\ref{fig:trc_1}, but for different assumptions on  the nature of the flow of matter after RLOF by the visual star. Columns to the right show examples of evolution without mass-loss via external Lagrange point   $\mathrm{L_2}$ (parameter $\delta_\mathrm{mt}=0$). 
Columns to the left show for clarity results of computations under assumptions made in
the present study: fraction of the matter lost by the donor and escaping via $\mathrm{L_2}$  is $\delta_\mathrm{mt}=0.1$, dimensionless specific angular momentum in the circumbinary torus  $\gamma_\mathrm{mt}=3$. The upper figures are for accepted in MESA standard value of critical accretion rate 
$\dot M_{cr}=\dot M_\mathrm{Edd}$; when this value is exceeded, isotropic wind from the compact object starts to blow (red line). The lower figures are for the case when critical accretion rate after which isotropic wind begins is set by us in MESA to 
$\dot M_{cr}=10^{-6}$ \ms yr$^{-1}$. It is seen that the effect of the parameters of non-conservative mass exchange 
$\delta_\mathrm{mt}$ and $\gamma_\mathrm{mt}$ is much substantial than that of the parameter  
$\beta_\mathrm{mt}=((dM/dt)_{\rm L1}-\dot M_{cr})/(dM/dt)_{\rm L1}$, 
which depends on the accepted value of $\dot M_{cr}$. Non-conservative outflow of the matter via the torus around the system leads to formation of a common envelope and cuts back the time length of the possible NULX stage.

\begin{figure}
\begin{center} 
\includegraphics[width=0.5\textwidth]{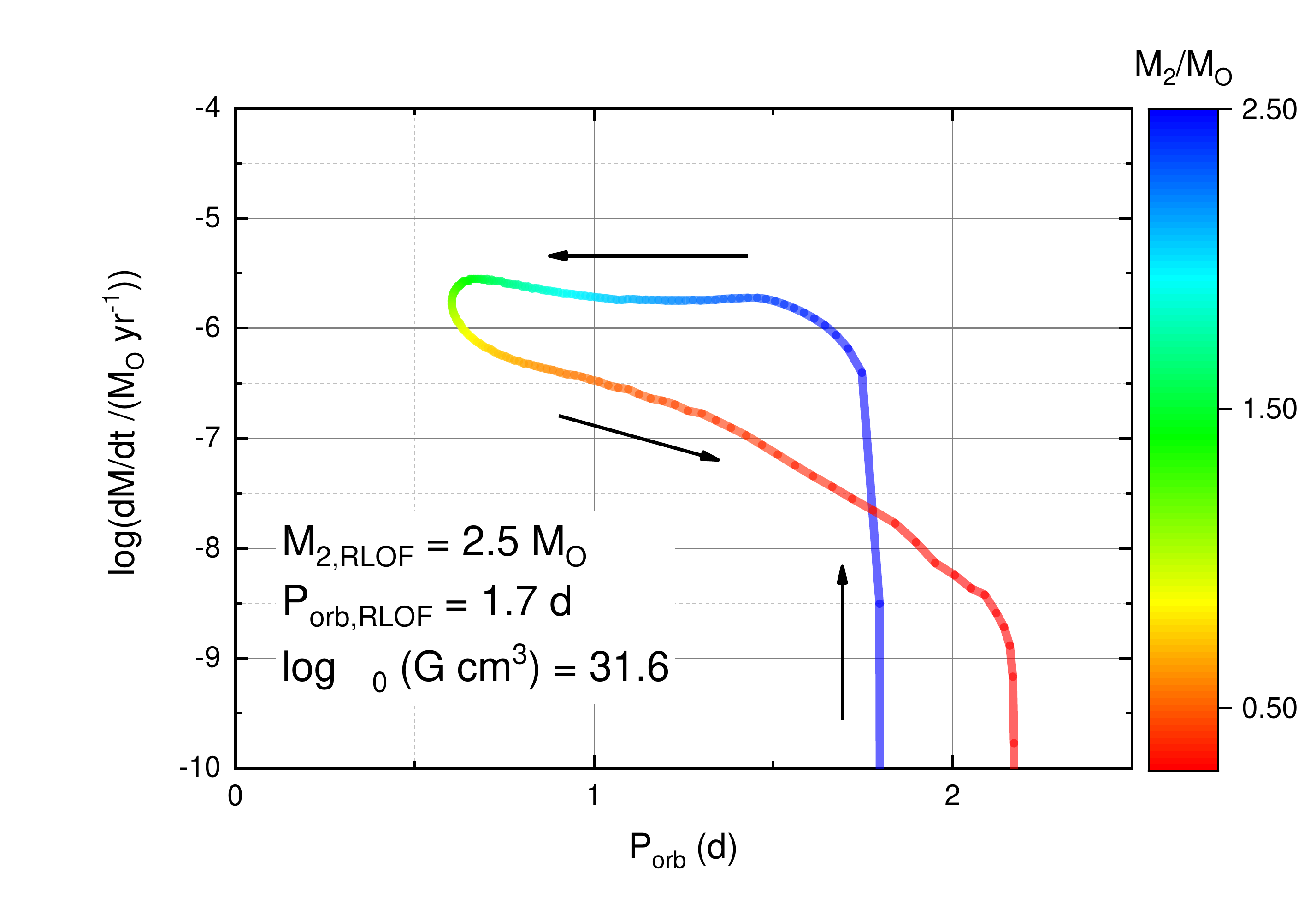}
\end{center}
\vfill
\includegraphics[width=0.5\textwidth]{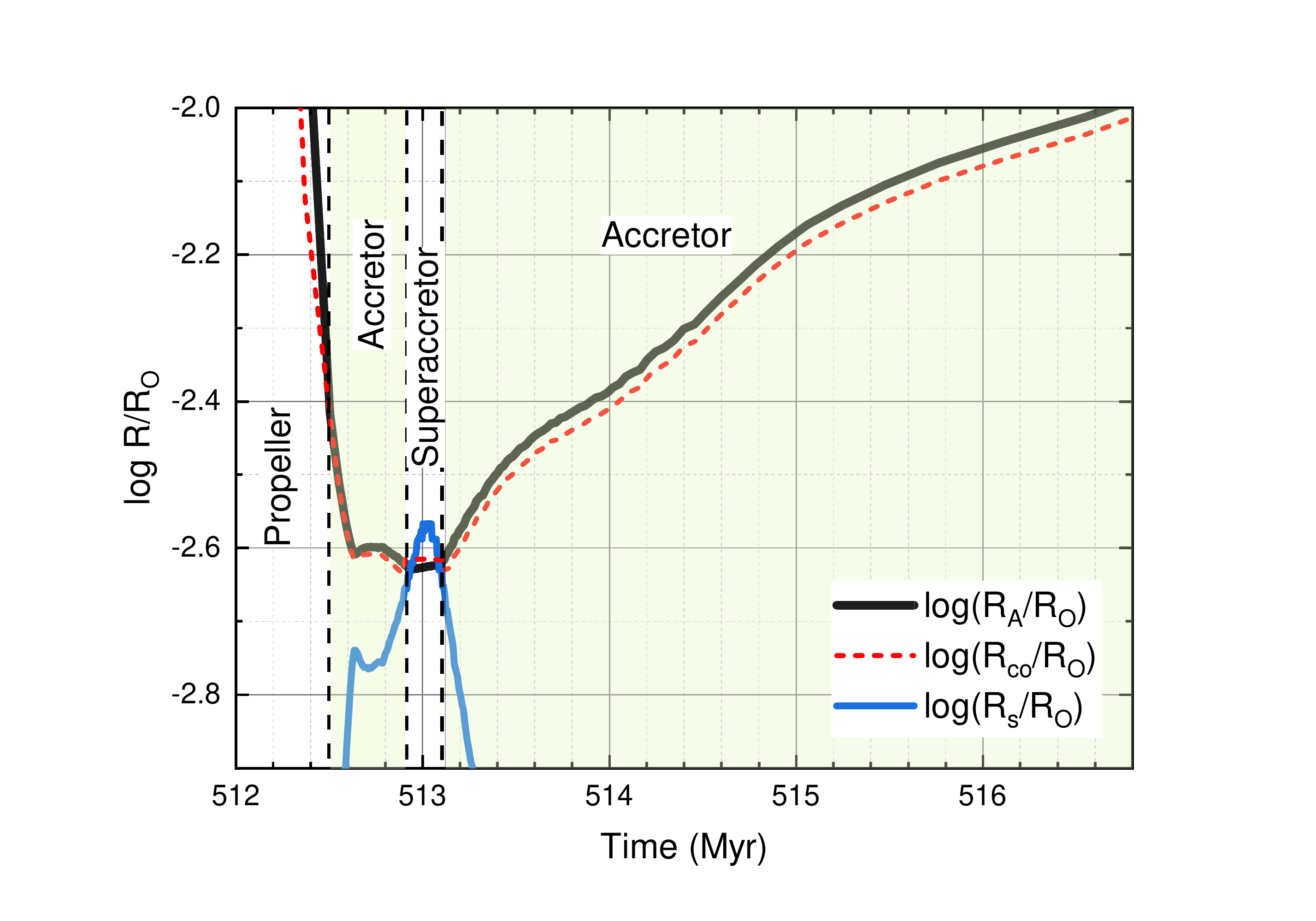}
\includegraphics[width=0.5\textwidth]{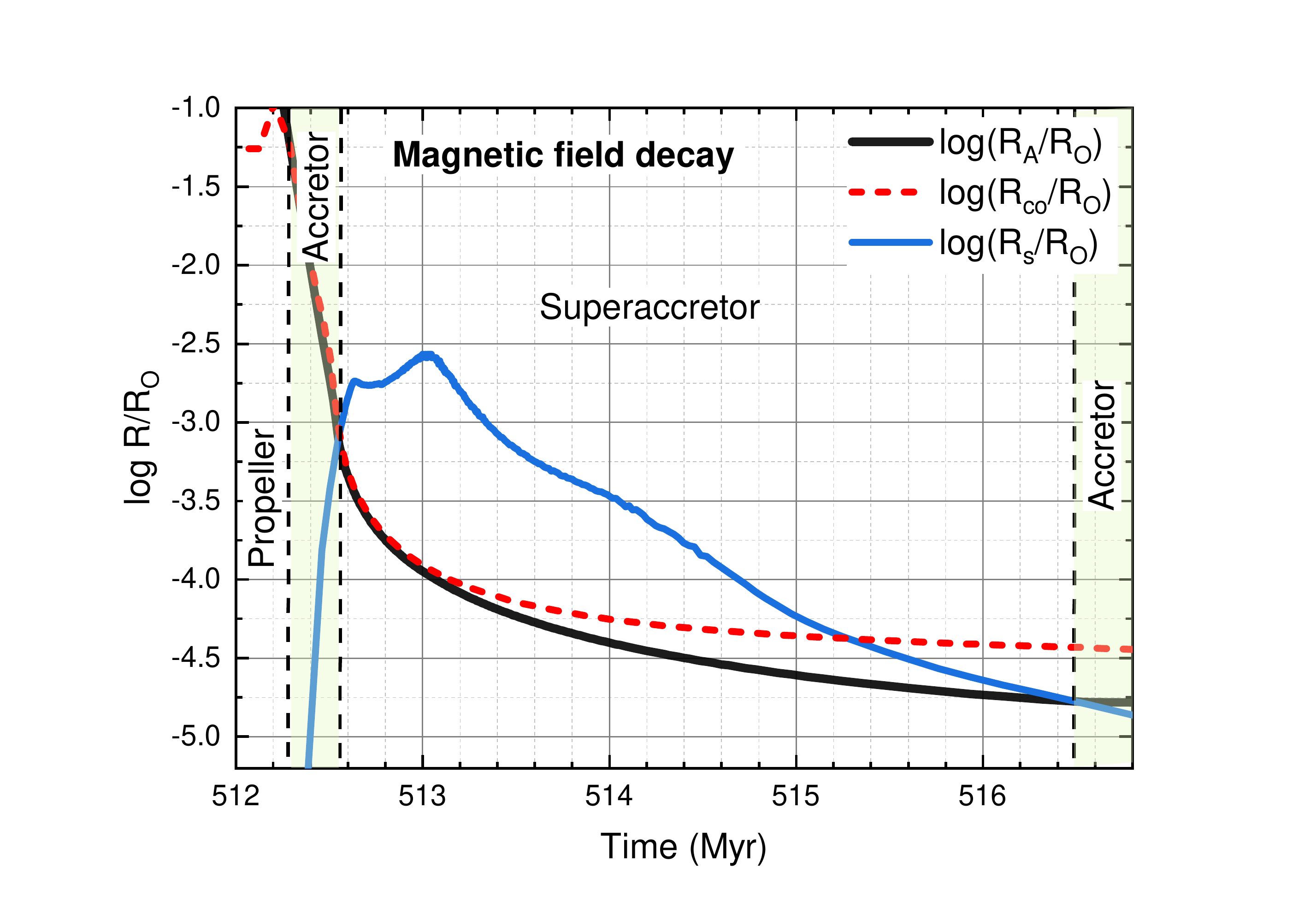}
\vfill
\includegraphics[width=0.5\textwidth]{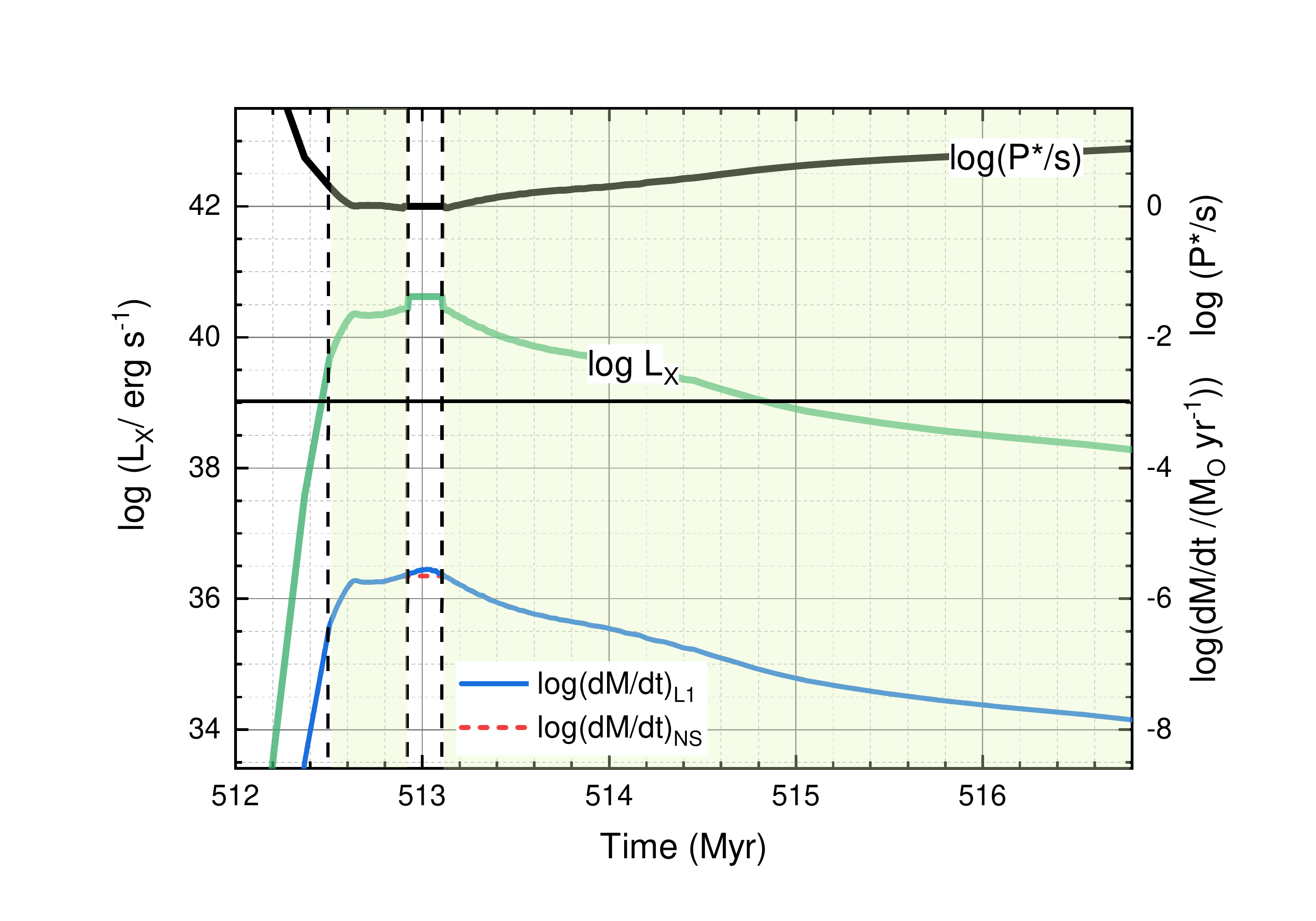}
\includegraphics[width=0.5\textwidth]{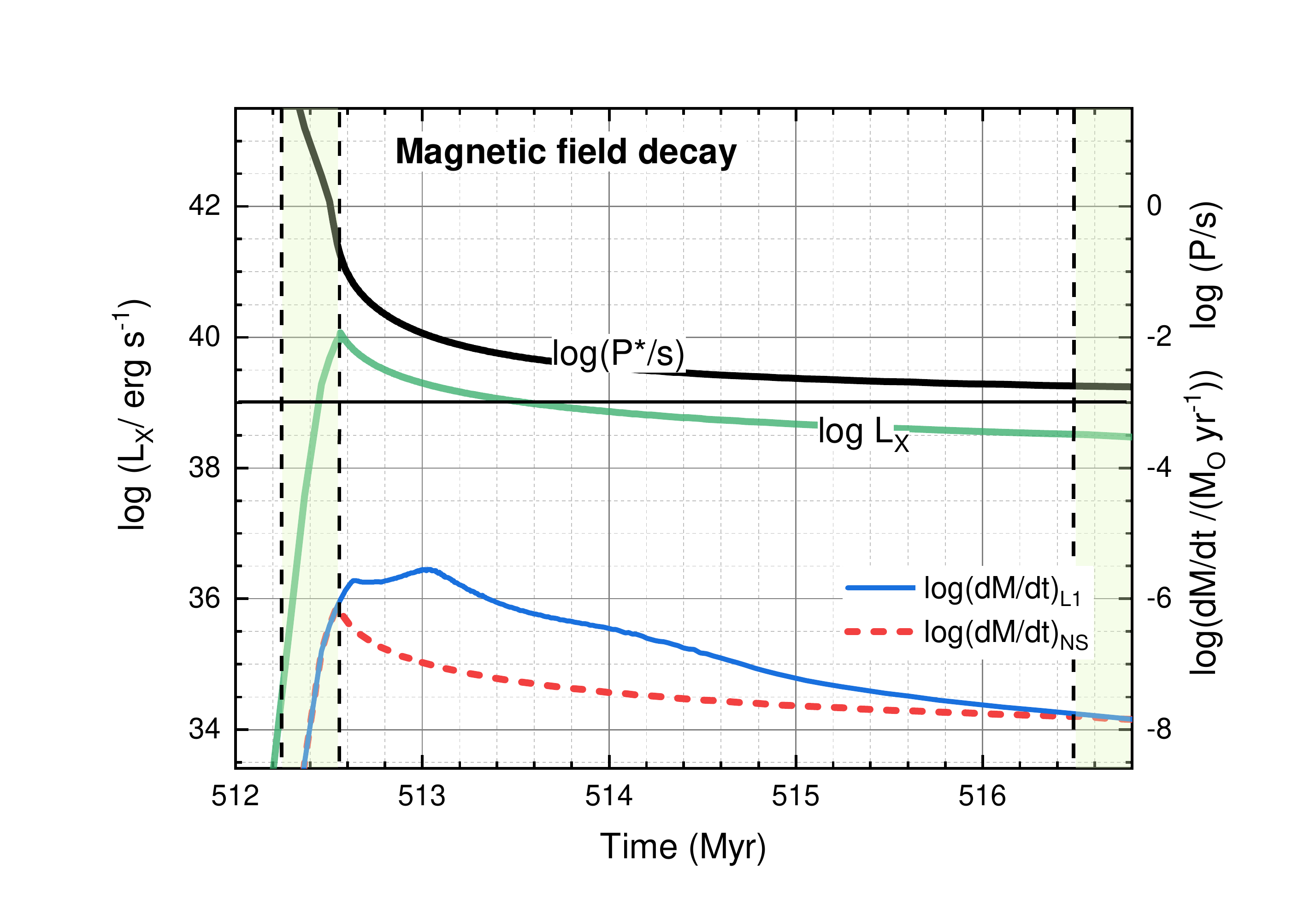}
\caption{An example of  evolutionary tracks at the stages of sub- and supercritical disc accretion onto NS. In the system  ``MS star ($M_2=2.5 M_\odot$) + NS'' orbital period at RLOF by visual component is  $P_\mathrm{orb}=2.75$~day. Panels to the left show the track without decay of the magnetic field. Panels to the right show the track with account of an exponential decay of magnetic field due to accretion.   
}
\label{fig:trc_2.5}
\end{figure}

\subsection*{An example of the track with a stable mass-transfer at the NULX stage}
\label{s:stable}
Figure \ref{fig:trc_2.5} shows an evolutionary track of the system with the mass of the visual star $M_2=2.5 M_\odot$ and orbital period at RLOF 
$P_\mathrm{orb}=2.75$~day. 
The same parameters as in Fig.~\ref{fig:trc_1} are shown.
At the stage of the stable mass-transfer NS is mainly in the state of accretion. The system does not enter  common envelope, orbital period of the system decreases to a minimum and then increases. Evolution of the system terminates by formation of a ``NS + helium white dwarf'' object. The stage of  NULX lasts for more than 
2~Myr. This time is long enough for the decay of the magnetic field of NS due to the accretion. Right panels of Fig.~\ref{fig:trc_2.5}  show the properties of the same system, but for the track computed with account of the decay of magnetic field that follows  Os{\l}owski et al. (2011) model:
$$
\mu(t)=(\mu_0-\mu_\mathrm{min})\times \exp\left({-\frac{\Delta M}{0.025\,M_\odot}}\right)+\mu_\mathrm{min},
$$
where  $\Delta M$ is the mass of accreted matter, 
$\mu_\mathrm{min}=10^{26}$ G~cm$^3$ -- the minimum magnetic momentum of NS.
It is clear that decay of the magnetic field results in the decrease of the radius of magnetosphere and respective decrease of 
$\dot M_\mathrm{cr}\sim R_A\sim \mu^{4/7}$ and X-ray luminosity. At this, the time length of the NULX stage also becomes two times shorter (bottom panel to the right). Note also that NS with decaying magnetic field stays predominantly in the supercritical accretion stage.  
Of course, due to the large uncertainty of the quantitative parameters in the problem of magnetic field decay, this example should be treated only as an indicator of the trend 
to reduction of luminosity and the number of NULX with low-mass visual components
($M_2\lesssim 3 M_\odot$).

{\bf REFERENCES}

\begin{enumerate}
\item M. Bachetti et al., \nat\ {\bf 514}, 202 (2014).
\item M. Basko and  R. Sunyaev, \mnras\ {\bf 175}, 395 (1976).
\item А. Belfiore et al., Nat. Astron.\ {\bf 4}, 147 (2020). 
\item M. Brightman et al., Nat. Astron.\ {\bf 2}, 312 (2018).
\item M. Brightman et al., \apj\ {\bf 895}, 127 (2020).
\item S. Carpano et al., \mnras\ {\bf 476}, L45 (2018).
\item A. Chashkina et al., \mnras\ {\bf 470}, 2799 (2017).
\item A. Chashkina et al., \aap\ {\bf 626}, A18 (2019).
\item H.-L. Chen et al., \mnras\ {\bf 445}, 1912 (2014).
\item A.~M~Cherepashchuk et al., \mnras\ {\bf 479}, 4844 (2018).
\item A.~M~Cherepashchuk et al., \mnras\ {\bf 485}, 2638 (2019).
\item A.~M~Cherepashchuk et al., New Astron. Rev. {\bf 89}, 101542 (2020).
\item R.~H.~D. Corbet et al., AAS/High Energy Astrophysics Division \#7 (2003).
\item A. P. Cowley and P.~C. Schmidtke, \aj\ {\bf 128}, 709 (2004).
\item M. de Kool, \apj\ {\bf 358}, 189 (1990).
\item H. P. Earnshaw et al., \mnras\ {\bf 483}, 5554 (2019).
\item А. Erdem and O. \"Otzt\"urk, \mnras\ {\bf 441}, 1166 (2014).
\item C.-A. Faucher-Gigu{\`e}re and V.M. Kaspi, \apj\ {\bf 643}, 332 (2006).
\item T. Fragos et al., \apjl\  {\bf 802}, L5 (2015).
\item N. Giacobbo and M. Mapelli, \mnras\ {\bf 480}, 2011 (2018).
\item J. Goliasch and L. Nelson, \apj\ {\bf 809} 80 (2015).
\item S. Grebenev, Astron. Lett.\ {\bf 43}, 464 (2017).
\item F. Gris\'{e} et al. \aap\ {\bf 486},  151 (2008).
\item M. Heida et al., \apjl\ {\bf 883}, L34 (2019).
\item G. Hobbs et al., \mnras\ {\bf 360}, 974 (2005).
\item J. Hurley et al., \mnras\ {\bf 329}, 897 (2002).
\item G. L. Israel et al., \mnras\ {\bf 466}, L48 (2017a).
\item G. L. Israel et al., Science {\bf 355}, 817 (2017b).
\item J.-A. Kennea et al., The Astronomer's Telegram {\bf 10809} (2017).
\item А. R. King, \mnras\ {\bf 393}, L41 (2009).
\item A. King, J.-P. Lasota, W. Kluzniak, \mnras\ {\bf 68}, 59 (2017). 
\item A. King and J.-P.~Lasota, \mnras\ {\bf 485}, 3588 (2019).
\item F. Koliopanos et al., \aap\ {\bf 608}, A47, (2017).
\item K. Kovlakas et al., \mnras\ {\bf 498}, 4790 (2020).
\item V.~M. Lipunov, \sovast\ {\bf 26}, 54 (1982).
\item V.~M. Lipunov, Astrophysics of neutron stars, New York: Springer (1992). [1987] 
\item V.~M.Lipunov et al.,   Astronomy Reports {\bf 53}, 915 (2009).
\item G.~V. Lipunova, Astron. Lett. {\bf 25}, 508 (1999).
\item K.~S.~Long and L.~P.~van Speybroeck, Accretion-Driven Stellar X-ray Sources, 
eds. W.H.G. Lewin and E.P.J. van den Heuvel, Cambridge, CUP, 119 (1983). 
\item A.-J. Loveridge et al., \apj\ {\bf 743}, 49 (2011).
\item P. Marchant et al., \aap\ {\bf 604}, A55 (2017).
\item M.~J.Middleton et al., \mnras\ {\bf 486}, 2 (2019).
\item S. Mineo et al., \mnras\ {\bf 419}, 2095 (2012).
\item D. Misra et al., arXiv:2004.01205 (2020).
\item S. Miyaji et al., PASJ {\bf 32}, 303 (1980).
\item С. Motch et al., \nat\ {\bf 514},  198 (2014).
\item R. Mushotzky et. al., \prtphs\ {\bf 155} 27 (2004).
\item A.~A. Mushtukov et. al., \mnras\  {\bf 454}, 2539 (2015).
\item A.~A. Mushtukov et. al., \mnras\  {\bf 467}, 1202 (2017).
\item A.~A. Mushtukov et. al., \mnras\  {\bf 486}, 2867, (2018).
\item S. Os{\l}owski et al., \mnras\ {\bf 413}, 461 (2011).
\item B. Paxton et al., \apjs\ {\bf 192}, 3 (2011).
\item K.~A. Postnov et al., IAU Symposium {\bf 346}, 219 (2019).
\item Y. Qiu et al., \apj\ {\bf 877}, 57 (2019).
\item G.~A.Rodr{\'\i}guez Castillo et al., \apj\ {\bf 895}, 60 (2020).
\item H. Sana et al., Science {\bf 337}, 444 (2012).
\item R. Sathyaprakash et al., \mnras\ {\bf 488}, L35 (2019).
\item S.~Y. Sazonov and I.~Khabibullin, \mnras\ {\bf 466}, 1019 (2017).
\item N.~I.~Shakura and R.~A.~Sunyaev,  \aap\ {\bf 500}, 33 (1973).
\item Y. Shao and  X.-D. Li, \apj\ {\bf 802}, 131 (2015).
\item L. Siess and  U. Lebreuilly, \aap\  {\bf 614}, A99 (2018).
\item G.~E.~Soberman et al., \aap\ {\bf 327}, 620 (1997).
\item L.~Tao et al., \apj\ {\bf 873}, 19 (2019).
\item S.~P. Trudolyubov et. al., \apj\ {\bf 663}, 487 (2007).
\item S. S. Tsygankov et al., \mnras\ {\bf 657}, 1101, (2016).
\item S. S. Tsygankov et al., .\aap\ {\bf 605}, A39 (2017).
\item G. Vasilopoulos et al., \mnras\ {\bf 491}, 4949 (2020).
\item J. Vink et al., \aap\ {\bf 369}, 574 (2001).
\item J. Vink, \aap\   {\bf 607}, L8 (2017).
\item I. Waisberg et al., \aap\ {\bf 623}, A47 (2019).
\item D.~J.~Walton et al., AN {\bf 332}, 354 (2011).
\item D.~J.~Walton et al., \apj\ {\bf 856}, 128 (2018).
\item R. F. Webbink, \apj\ {\bf 277}, 355 (1984).
\item G. Wiktorowicz et al.,  \apj\ {\bf 846}, 17 (2017).
\item G. Wiktorowicz et al.,  \apj\ {\bf 875}, 53 (2019).
\item S.~Yu and C.~S.~Jeffery, \aap\ {\bf 521}, A85 (2010).
\item Y. Zhang et al., \apj\ {\bf 879}, 61 (2019).
\end{enumerate}
\label{lastpage}

\end{document}